\documentclass [12pt]{article}
\usepackage{amsfonts,amsmath,amssymb}
\usepackage{graphicx}

\newfont{\twelvemsb}{msbm10 scaled\magstep1}
\newfont{\eightmsb}{msbm8}
\newfam\msbfam
\textfont\msbfam=\twelvemsb \scriptfont\msbfam=\eightmsb
\catcode`\@=11
\def\Bbb{\ifmmode\let\next\Bbb@\else
\def\next{\errmessage{Use \string\Bbb\space only in math mode}}\fi\next}
\def\Bbb@#1{{\fam\msbfam{{#1}}}}

\newcommand{\be}{\begin{equation}}
\newcommand{\ee}{\end{equation}}
\newcommand{\ba}{\begin{eqnarray}}
\newcommand{\ea}{\end{eqnarray}}
\begin{document}
\sloppy
\renewcommand{\thefootnote}{\fnsymbol{footnote}}
\newpage
\setcounter{page}{1} \vspace{0.7cm}
\begin{flushright}
10/08/12
\end{flushright}
\vspace*{1cm}
\begin{center}
{\bf TBA-like equations and Casimir effect in (non-)perturbative AdS/CFT}\\
\vspace{1.8cm} {\large Davide Fioravanti $^a$ and
Marco Rossi $^b$
\footnote{E-mail: fioravanti@bo.infn.it, rossi@cs.infn.it.}}\\
\vspace{.5cm} $^a$ {\em Sezione INFN di Bologna, Dipartimento di Fisica, Universit\`a di Bologna, \\
Via Irnerio 46, Bologna, Italy} \\
\vspace{.3cm} $^b${\em Dipartimento di Fisica dell'Universit\`a
della Calabria and INFN, Gruppo collegato di Cosenza, I-87036
Arcavacata di Rende, Cosenza, Italy} \\
\end{center}
\renewcommand{\thefootnote}{\arabic{footnote}}
\setcounter{footnote}{0}
\begin{abstract}
{\noindent We consider high spin, $s$, long twist, $L$, planar operators (asymptotic Bethe Ansatz) of strong ${\cal N}=4$ SYM. Precisely, we compute the minimal anomalous dimensions for large 't Hooft coupling $\lambda$ to the lowest order of the (string) scaling variable $\ell \sim L/  (\ln \mathcal{S} \sqrt{\lambda})$ with GKP string size $\sim\ln  \mathcal{S}\equiv 2 \ln (s/\sqrt{\lambda})$. At the leading order $(\ln \mathcal{S}) \cdot \ell ^2 $, we can confirm the $O(6)$ non-linear sigma model description for this bulk term, without boundary term $(\ln \mathcal{S})^0$. Going further, we derive, extending the $O(6)$ regime, the exact effect of the size finiteness. In particular, we compute, at all loops, the first Casimir correction $\ell ^0/\ln \mathcal{S}$ (in terms of the infinite size $O(6)$ NLSM), which reveals only one massless mode (out of five), as predictable once the $O(6)$ description has been extended. Consequently, upon comparing with string theory expansion, at one loop our findings agree for large twist, while reveal for negligible twist, already at this order, the appearance of wrapping. At two loops, as well as for next loops and orders, we can produce predictions, which may guide future string computations.}
\end{abstract}
\vspace{5cm}
\newpage
\section{General setting, aims and results}
\setcounter{equation}{0}

The investigations, tests and deeper understanding of the AdS/CFT correspondence \cite{M, GKP, W} were greatly boosted by the discovery of integrability in planar ${\cal N}=4$ SYM  \cite {MZ,BS}. As for integrability, its current form states that the spectrum of anomalous dimensions of (composite) trace operators may be found upon solving a set of Thermodynamic Bethe Ansatz (TBA) integral equations \cite {TBA}. For long operators the quantisation conditions for rapidities in the TBA approach reduce to algebraic Asymptotic Bethe Ansatz (ABA) equations (S-matrix quantisation conditions), as long as wrapping corrections \cite{WRA} are negligible.

Among the most interesting operators there are the (twist) Wilson operators and the simplest examples and paradigm of these are the scalar operators of ${\cal N}=4$ SYM. Closed under renormalisation, the $sl(2)$ scalar operators make use of only one (out of three) complex scalar ${\cal Z}$ and the (light-cone) covariant derivative ${\cal D}$ and enjoy the sketchy form
\begin{equation}
{\mbox {Tr}} ({\cal D}^s {\cal Z}^L)+.... \, , \label {sl2op}
\end{equation}
where dots stand for permutations. This composite single trace operator has of course Lorentz spin $s$ and twist (or length) $L$, with minimum value $L=2$ for which (a descendant\footnote{In this case, the spin may be shifted by a finite amount which does not affect our analysis and results at high spin.} of) the GKP 'vacuum' solution is realised \cite{Gubser:2002tv}. In general, the AdS/CFT correspondence relates operators (\ref {sl2op}) to spinning folded closed strings on $\text{AdS}_5\times\text{S}^5$ spacetime, with $\text{AdS}_5$ and $\text{S}^5$ angular momenta $s$ and $L$, respectively, the 't Hooft coupling $\lambda $ being connected to the string tension $T=\frac{\sqrt{\lambda}}{2\pi}$ \cite{Gubser:2002tv, FT}.

Actually, in the semiclassical string (world-sheet) expansion one starts, as usual, from the classical action at large tension $\propto \sqrt{\lambda}$ and uses the (finite) classical $\text{AdS}_5$ angular momentum $\mathcal{S}=\frac{s}{\sqrt{\lambda}}$ and the (finite) classical $\text{S}^5$ angular momentum  $\mathcal{J}=\frac{L}{\sqrt{\lambda}}$. Therefore, the strong coupling limit comes together with the high spin, large twist limit, in the double scaling form
\begin{equation}
g \rightarrow \infty \, , \quad  s\rightarrow \infty \, , \quad L\rightarrow \infty \, , \quad \mathcal{S}=\frac{s}{\sqrt{\lambda}}={\mbox {fixed}} \, ,
\quad \mathcal{J}=\frac{L}{\sqrt{\lambda}}={\mbox {fixed}} \label {Slimit} \, ,
\end{equation}
with the parametrisation $\lambda =8\pi ^2 g^2$. This scaling limit is not so natural, neither usual when dealing with the gauge theory where a perturbative loop expansion in (positive integers) powers of $\lambda$ takes place at finite $L$. Of course integrability helps and we showed in \cite{FGR5} how the minimal anomalous dimension of (\ref {sl2op}) enjoys the high spin expansion (at fixed, but generic $g$ and $L$)
\be
\gamma(g,s,L)= f(g) \ln s + f_{sl}(g,L) + \sum_{n=1}^\infty \gamma^{(n)}(g,L)  \, (\ln s)^{-n} + O\left ( (\ln s)/s \right ) \, , \label {subl}
\ee
as a massless series of large size expansion in the size parameter $\ln s$. Impressively, the coefficient of the leading high spin $(\ln s)$ term (fixed $L$)\footnote{This is the so-called {\it universal scaling function}, $f(g)$, which does not depend on $L$ and equals twice the {\it cusp anomalous dimension} of a light-like Wilson loop \cite{KM}.} was obtained from the solution of a linear integral equation directly derived from the ABA via the root density approach \cite {BES}. Besides, the sub-leading $(\ln s)^0$ contribution $f_{sl}(g,L)$ received already much attention in \cite
{FRS} by a Non-Linear Integral Equation (NLIE) and in \cite{BFR} by a linear integral equation (LIE). Explicit weak and strong coupling expansions are present in \cite {FZ}  by solving the LIE and agree with string theory computations \cite {BFTT}. Importantly, it is believed that both $f(g)$ and $f_{sl}(g,L)$ are exactly given by this approach based on the ABA without wrapping. In any case, thanks to the integrability in the form of ABA equations or, practically, to the exact expressions given by (N)LIEs for all the coefficients in (\ref{subl}), the strong coupling limit may be considered, but this would happen {\it after} the high spin expansion. Furthermore, at fixed and generic $g$ ($s\rightarrow +\infty$) also the long operators
\be
L\rightarrow \infty \, , \quad j=\frac {L-2}{\ln s}  \, \quad {\mbox {fixed}}
\label {jlimit}
\ee
can be considered to give rise to a {\it generalised scaling function} $f(g,j)$ at leading order $\gamma=f(g,j)\ln s+\dots$ \cite{BGK,SS}. Similarly, they enjoy the expansion \cite{FIR,FGR5}
\be
\gamma (g,s,j)=f(g,j)\ln s+ \sum _{n
=0}^{\infty} f^{(n)}(g,j) (\ln s )^{-n} + O\left ( (\ln s )/s \right ) \, , \label {gamma-j}
\ee
generating an infinite number of other scaling functions. Yet, in the aforementioned semiclassical string (world-sheet) expansion the natural parameter is not even $j$, but of course its scaled version \cite{BGK,SS,FTT,AM}
\be
\ell_s=\pi\frac{\mathcal{J}}{ \ln \mathcal{S}} =\frac{\pi L}{ \sqrt{\lambda} \ln \mathcal{S}}\simeq \pi \frac{j}{ \sqrt{\lambda} }\, ,
\label {ell-string}
\ee
and analogously to (\ref{gamma-j}) we can expect and will compute in the following the string expansion
\be
\gamma(g,\mathcal{S},\mathcal{J})= f_s(g, \ell_s) \ln \mathcal{S} + \sum_{n=0}^\infty f_s^{(n)}(g,\ell_s)  \, (\ln \mathcal{S})^{-n} + O\left ( (\ln \mathcal{S})/ \mathcal{S} \right ) \, , \label {gamma-string}
\ee
where the index $s$ stands for 'string'. Despite these changes of variables going from the gauge theory (weak coupling) to the string theory (strong coupling) it is predictable that the two expansions (\ref{gamma-j}) and (\ref{gamma-string}) do not match (once expressed in the same set of variables). For the order of limits does matter and adds hurdles to check and understand the correspondence, whose nature is already problematic because of its strong/weak coupling character. Once again, integrability will help us, as we will see in the following, and show how it is a fundamental tool to understand quantitatively the physics of the order of limits in strong/weak dualities.

In this paper we wish to compute the anomalous dimension, $\gamma(g,\mathcal{S},\mathcal{J})$ in the form (\ref{gamma-string}), {\it i.e.} in 'the string limit' (\ref {Slimit}), starting from the ABA equations. In particular, we will pay attention to the Casimir term $f_s^{(1)}(g,\ell_s)$, which does not vanish at first order $\ell_s^0$, in manifest 'disagreement' with the 'gauge limit' expansions (\ref{gamma-j}) and (\ref{subl}): $f^{(1)}_0(g)\equiv 0$ and $\gamma^{(1)}(g,L)\equiv 0$ with $f^{(n)}(g,j)=\sum^{\infty}_{r=0} f_r^{(n)}(g) j^r$ \cite{FGR5}. On the contrary, we will show that the leading part of this first finite size correction reveals, in the string regime (\ref {Slimit}), the number of degrees of freedom for the model, as it realises the central charge in the usual conformal limit \cite{Bloete:1986qm}. Additionally, the higher logarithmic inverse powers are associated to irrelevant perturbations of the $O(6)$ NLSM.

In more detail, we will move from the ABA equations \cite{BS} without including wrapping corrections \cite{BES}. Since we want to analyse the string regime (\ref {Slimit}), we find it convenient and efficient to equivalently rewrite the ABA equations as one single NLIE and discard the root non-linear contribution $\sim 1/s^2$. We end up with a purely {\it hole NLIE} which captures the perturbative and non-perturbative regimes of string energy (up to $\sim 1/s^2$). In particular, we will develop and apply it further in the non-perturbative regime of small $\ell_s \ll 1$\footnote{This limit gives the lowest order in $\ell_s$ of string perturbation theory. The latter could usefully be explored by a different development of the hole NLIE suitable to gain higher orders of $\ell_s$.} or $j\ll \sqrt{\lambda}$, where low energy and momentum of the scalar excitations in (\ref{sl2op}) are both comparable with their exponentially small (rest) mass $m\sim e^{-\frac{\sqrt{\lambda}}{4}}$ \footnote{After looking at the energy dispersion relation $E^2=m^2+p^2+ O(p^4)$ as a function of the momentum $p$, we realise easily that for small $p$ (comparable with $m$ or smaller) we are in the aforementioned non-perturbative regime, which is also the relativistic one. Otherwise, as the momentum $p$ grows, the $O(p^4)$ starts to contribute significantly. Once the momentum is much bigger than the (non-perturbative in $1/\sqrt{\lambda}$) mass-gap, the latter can be neglected: hence, this case becomes tractable by string perturbation theory, but the excitations are no longer relativistic . A deep and precise illustration of the dispersion relations may be found in \cite {BAS}.}. This new route highlights that these excitations scatter as the particles of the $O(6)$ NLSM in the thermodynamic, {\it i.e.} infinite size, limit, and the scaled R-charge $\ell_s$ serves as magnetic field. For this reason, the 'bulk' energy, -- i.e. the leading part $\propto$(size)$\sim \ln \mathcal{S}$ --,  of the $\text{AdS}_5\times\text{S}^5$ sigma model, after subtracting the cusp contribution, is given by that of its scalar part, {\it i.e.} an $O(6)$ Non-Linear Sigma Model (NLSM) as proposed by \cite {AM} (cf. below for more details). Yet, we will go beyond the thermodynamic limit by computing exactly the energy for finite size up to exponentially small $\sim e^{-\ln \mathcal{S}}=1/\mathcal{S}$ corrections. In fact, as in the 'gauge order of limit' (\ref{subl},\ref{jlimit},\ref{gamma-j}) \cite{FGR5}, the expansion shows to be of massless type, i.e. in terms of (positive integer) powers of $1/($size$)\sim 1/ \ln \mathcal{S}\sim 1/ \ln s$  \footnote{Still, the attentive reader should have noticed that $\ell_s$ contains $1/ \ln \mathcal{S}$, giving rise to a complicate expansion.}. Albeit a detailed account of the results may be found in the concluding Section 7, we can here anticipate the agreement with the one-loop string result \cite{GRRT} for ${\mathcal{J}}\neq 0$ of the Casimir energy, namely the first finite size correction ($\sim 1/ \ln \mathcal{S}$). The discrepancy for ${\mathcal{J}}= 0$  manifests how wrapping enters already at this order by adding four massless (scalar) degrees of freedom to the 'central charge': thus, their contribution should be dumped as $\sim e^{-(const.) \mathcal{J}}$ for large ${\mathcal{J}}$ and {\it a posteriori} their mass may be deduced to be $\sim {\mathcal{J}}/\ln \mathcal{S}$   \footnote{This mechanism is quite clear also in the algebraic curve approach \cite{GSSV}.}. On the contrary, in \cite{nlie-reciprocity} we argue how for 
small $\lambda$ the wrapping needs to correct, as first order, the term $\sim 1/s^2$. Finally, some predictions at subsequent orders along with the effective possibility to expand exact formul{\ae} should help string perturbative regularisations and computations.

The paper is organised as follows. In Section 2 we write the single NLIE describing the ground state
for generic values of root number (spin), length (twist) and coupling. In Sections 3 and 4 we perform the scaling limit
(\ref {Slimit}) on the NLIE for the density and on the non-linear functional for the anomalous dimension. In Section 5 we rewrite the NLIE for the density and the non-linear functional giving the anomalous dimension in the form of a finite size perturbation of the infinite space $O(6)$ NLSM\footnote{This model is not the kink (or massless) limit of the finite size $O(6)$ NLSM as resulting, {\it e.g.}, from TBA. The missing of four massless modes in the 'central charge' is a clear manifestation of this fact.}. In Section 6 we discuss and find the finite size correction $\sim \frac{1}{\ln \mathcal{S}}$ for anomalous dimension. Results, conclusions and open problems are summarised in Section 7.

\section{From many Bethe equations to a single nonlinear integral equation}
\setcounter{equation}{0}

Let us focus on the scalar twist sector (\ref {sl2op}) and on its set of ABA equations \cite {BS,BES}.
Using notations that will be useful in the following, we write the ABA equations in their logarithmic form
\be
Z(u_k)=2\pi n_k \, , \quad n_k \in \mathbb {Z} \, , \quad k=1,...,s \, . \label {Z-eq}
\ee
In (\ref {Z-eq}) the counting function $Z(u)$ is
\be
Z(u)=\Phi (u)-\sum _{k=1}^{s}
\phi (u,u_k) \, , \label {Z}
\ee
where
\be
\Phi (u)=\Phi _0(u)+\Phi _H (u) \, , \quad \phi (u,v)=\phi _0(u-v)+\phi _H (u,v) \, , \nonumber
\ee
and
\ba
\Phi _0(u) &=&-2L \arctan 2u \, , \quad \Phi _H(u)=-iL \ln \left ( \frac {1+\frac {g^2}{2{x^-(u)}^2}}{1+\frac {g^2}{2{x^+(u)}^2}} \right )\, ,  \label {Phi} \\
\phi _0(u-v)&=& 2\arctan (u-v) \, ,\quad \phi _H(u,v)=-2i \left [ \ln \left ( \frac {1-\frac {g^2}{2x^+(u)x^-(v)} }{1-\frac {g^2}{2x^-(u)x^+(v)}} \right )+i\theta (u,v)\right] \, . \label {phi}
\ea
In (\ref {Phi}, \ref {phi}) the usual notations
\begin{equation}
x^{\pm}(u_k)=x(u_k\pm i/2) \, , \quad x(u)=\frac
{u}{2}\left [ 1+{ \sqrt {1-\frac {2g^2}{u^2}}} \right ] \, , \quad
\lambda =8\pi ^2 g^2 \, ,
\end{equation}
$\lambda $ being the 't Hooft coupling, are used.
In addition, $\theta (u,v)$ is the 'dressing factor'
which can be written \cite {BES} as
\begin{equation}
\theta (u,v)=\sum _{r=2}^{\infty}\sum _{\nu =0}^{\infty} g^{2r+2\nu-2}2^{1-r-\nu}c_{r,r+1+2\nu}(g)
[q_r(u)q_{r+1+2\nu}(v)-q_r(v)q_{r+1+2\nu}(u)] \, ,
\end{equation}
where \cite {BES,FRj}
\be
c_{r,r+1+2\nu}(g)= 2 (-1)^{\nu} (r-1) (r+2\nu) Z_{r-1,r+2\nu}(g) \, , \quad Z_{n,m}(g)=\int _{0}^{\infty} dt \frac{J_{n}(\sqrt{2}gt) J_{m}(\sqrt{2}gt)}{t(e^{t}-1)} \, , \label {cZ}
\ee
with $J_n(t)$ the usual Bessel function of the first kind.
Finally,
\begin{equation}
q_r(u)=\frac {i}{r-1} \left [ \left (\frac {1}{x^+(u)}\right
)^{r-1}-\left (\frac {1}{x^-(u)}\right )^{r-1} \right ] \, ,
\end{equation}
is the expression of the $r$-th charge in terms of the rapidity $u$.
In particular, the (asymptotic) anomalous dimension $\gamma $ of (\ref {sl2op}) is given by the eigenvalues of the second charge:
\be
\gamma =g^2 \sum _{k=1}^s q_2(u_k) \, .
\ee
It is widely known \cite{BGK, FRS,BFR} that $L+s$ real points $\upsilon _k$ satisfy condition (\ref {Z-eq}): $s$ of them are the genuine Bethe roots $u_k$, while the remaining $L$ points are 'spurious' solutions called 'holes' (and denoted as $x_h$). For the ground state (lowest energy,
{\it i.e.} lowest anomalous dimension) both the Bethe roots and holes are symmetrically distributed around the origin. Besides, one should distinguish between $L-2$ small or internal holes $x_h$, $h=1,...,L-2$, lying in the interval $[-c,c]$, and two large holes, $x_{L-1}=-x_L$. Bethe roots lie in the two regions $[-b,-c]$, $[c,b]$, with $0<b<x_L$. Concentrating on the ground state, we wish to convert (\ref {Z}) into an integral equation so that we can better perform and control a large parameter regime. To this aim, we can apply the general strategy of \cite {FMQR} to deal with real roots and holes and move on from the simple formula
\be
\sum _{k=1}^{s} O(u_k)+\sum _{h=1}^{L} O(x_h)=-\int _{-\infty}^{+\infty} \frac {dv}{2\pi} O(v)
\frac{d}{dv}Z (v)+  \int _{-\infty}^{+\infty}\frac {dv}{\pi}
O(v)\frac {d}{dv} {\mbox {Im}} \ln
[1+(-1)^{L} \, e^{iZ(v-i0^+)}] \, . \label {cauchy}
\ee
We eventually obtain for the Fourier transform\footnote {We define the Fourier transform of $f(u)$ as $\hat f(k)=\int \limits _{-\infty}^{+\infty} du \ e^{-iku} f(u)$.} $\hat \sigma (k)$ of the 'density' $\sigma (u)= \frac{d}{du} Z(u)$ the nonlinear integral equation
(due to parity properties $\hat \sigma (k)=\hat \sigma (-k)$ we restrict to the region $k>0$):
\ba
\hat \sigma (k)&=& \frac{\pi L}{\sinh \frac{k}{2}}[e^{-\frac{k}{2}}-J_0(\sqrt{2} gk) ]
+ \frac{2\pi e^{-k}}{1-e^{-k}} P(k;\{x_h\}) - \frac{2ik e^{-k}}{1-e^{-k}}\hat L(k)  \label {sigmaeq} \\
&-& \frac{g^2 \pi k}{\sinh \frac{k}{2}} \int _{0}^{+\infty} dt
 \mathcal{\hat K} ( \sqrt{2} gk, \sqrt{2} gt ) \Bigl [ 2P(t;\{x_h\}) e^{-\frac{t}{2}}+2L e^{-\frac{t}{2}}- \frac{it \hat L(t)(1-e^{-t})}{\pi \sinh \frac{t}{2}}+ \frac{e^{-\frac{t}{2}}}{\pi}\hat \sigma (t) \Bigr ] \, , \quad k>0 \, .\nonumber
\ea
In (\ref {sigmaeq}) we have used the shorthand notations
\be
P(t;\{x_h\})=\sum _{h=1}^L \left (\cos t x_h -1 \right ) \, , \quad L(u)= {\mbox {Im}} \ln
[1+(-1)^{L} \, e^{iZ(u-i0^+)}] \, ,
\ee
while the 'BES' kernel
\be
\mathcal{\hat K} (t,t')=\frac{2}{tt'}\left [ \sum _{n=1}^{\infty}n J_n (t) J_n (t') + 2 \sum _{k=1}^{\infty} \sum _{l=0}^{\infty} (-1) ^{k+l}c_{2k+1,2l+2}(g) J_{2k}(t) J_{2l+1}(t') \right ] \, , \label {magic}
\ee
is defined in \cite {ES,BES}.

Equation (\ref {sigmaeq}) should be used together with equation
\ba
S(k)&=& \frac{L}{k}[1-J_0(\sqrt{2} gk) ] -g^2 \int _{0}^{+\infty} dt
 \mathcal{\hat K} ( \sqrt{2} gk, \sqrt{2} gt ) \frac{P(t;\{x_h\})-L(e^{-\frac{t}{2}}-1)-\frac{it}{\pi}\hat L(t) }{\sinh \frac{t}{2}} - \nonumber \\
&-& g^2 \int _{0}^{+\infty} dt
\mathcal{\hat K} ( \sqrt{2} gk, \sqrt{2} gt ) e^{-\frac{t}{2}} \frac{t}{\sinh \frac{t}{2}} S(t)  \, , \quad k>0 \, , \label {Seq}
\ea
for the even function
\be
S(k)=\frac {\sinh \frac {|k|}{2}}{\pi |k|} \Bigl \{ \hat \sigma (k) + \frac{2ik e^{-|k|}}{1-e^{-|k|}}\hat L(k)+ \frac{\pi L}{\sinh \frac{|k|}{2}} \left (1-e^{-\frac{|k|}{2}} \right )
- \frac {2\pi e^{-|k|}}{1-e^{-|k|}}  \sum _{h=1}^{L} \left[ \cos kx_h- 1 \right] \Bigr \} \label {Sdef} \, ,
\ee
which is related to the anomalous dimension by the simple relation\footnote{We remark that formula (\ref {S-gamma}) is exact for any values of the spin $s$ and is proven in \cite{nlie-reciprocity}. It extends to generic values of the spin the relation $\gamma = \frac{1}{\pi}\hat \sigma (0) \Bigl|_{higher \ loops}$ which holds \cite {KL} only in the limit $s\rightarrow +\infty$. In \cite{BFR} we have already extended it to the subleading term $(\ln s)^0$.}
\be
\gamma =2 S(0) \, . \label {S-gamma}
\ee
In next section we will study equations (\ref {sigmaeq}, \ref {Seq}) in the limit (\ref {Slimit}), in which the spin $s$, the twist $L$ and the coupling constant $g$ go to infinity.

\section{High spin/large coupling expansion: the hole NLIE}
\setcounter{equation}{0}

We consider the equation for $\hat \sigma (k)$ and consider the high spin regime ({\it cf.} for instance \cite {ES,BES,BBKS,BKK,FZ,FIR,BDM,FGR3,FGR5}), bearing in mind that we wish also to perform a large $g$ expansion in the string scaling (\ref{Slimit}).
When $s\rightarrow \infty$ the equation for $\hat \sigma (k)$ becomes linear\footnote{In this perspective, the holes are supposed to be given or determined otherwise, as we want to focus here our attention on the roots. In the next Section, we will abandon this view and notice that the holes give rise to a non-linear integral equation.}, since the nonlinear terms contained in it are approximated by a constant:
\be
\int _{-\infty}^{\infty} \frac{dk}{2\pi} e^{iku} \frac{2ik e^{-|k|}}{1-e^{-|k|}} \hat {L}(k)=
2\ln 2 + O\left ( \frac{1}{s^2} \right ) \, . \label {non-lin}
\ee
This property has been widely used in the previously quoted literature related to the high spin limit (as for the one loop case {\it cf.} \cite{FRS}). We provide an analytic proof in \cite{nlie-reciprocity}.

By means of (\ref {non-lin}) we arrive at the following integral equation:
\ba
\hat \sigma (k)&=& \frac{\pi L}{\sinh \frac{k}{2}}[e^{-\frac{k}{2}}-J_0(\sqrt{2} gk) ]
+ \frac{2\pi e^{-k}}{1-e^{-k}} P(k;\{x_h\}) - (2\ln 2) \, 2\pi \delta (k)- \label {sigmaeq2} \\
&-& \frac{g^2 \pi k}{\sinh \frac{k}{2}} \int _{0}^{+\infty} dt
 \mathcal{\hat K} ( \sqrt{2} gk, \sqrt{2} gt ) \Bigl [ 2P(t;\{x_h\}) e^{-\frac{t}{2}}+2L e^{-\frac{t}{2}}+ \frac{e^{-\frac{t}{2}}}{\pi}\hat \sigma (t) \Bigr ] + O\left ( \frac{1}{s^2} \right ) \, , \quad k>0 \, . \nonumber
\ea
We can call it {\it hole NLIE} as non-linearity is confined in the hole term $P(t;\{x_h\})$ by using the strategy of converting it into a non-linear integral as in Section 2 \cite{FMQR} (cf. also beginning of Section 5). Moreover, we can recognise the splitting of its solution into
\be
\hat \sigma (k)= (L-2) \hat \sigma ^{(1)}(k) + \hat \sigma (k)|_{holes} \, , \label {first-split}
\ee
where $\hat \sigma ^{(1)}(k)$ is known, as being the Fourier transform of the density corresponding to the first generalised scaling function (i.e. the part of the density proportional to $\ln s \cdot \frac{(L-2)}{\ln s}$: see the first of \cite {FGR1} for details),
\be
\hat \sigma ^{(1)}(k)=\frac{\pi }{\sinh \frac{k}{2}}[e^{-\frac{k}{2}}-J_0(\sqrt{2} gk) ]
-\frac{g^2 \pi k}{\sinh \frac{k}{2}} \int _{0}^{+\infty} dt
 \mathcal{\hat K} ( \sqrt{2} gk, \sqrt{2} gt ) \Bigl [ 2 e^{-\frac{t}{2}}+ \frac{e^{-\frac{t}{2}}}{\pi}\hat \sigma ^{(1)}(t) \Bigr ] \, , \quad k>0 \, , \label {sigma(1)}
\ee
and $\hat \sigma (k)|_{holes}$ takes full account of the dependence on the hole positions via the equation
\ba
\hat \sigma (k)|_{holes}&=& \frac{2\pi}{\sinh \frac{k}{2}}[e^{-\frac{k}{2}}-J_0(\sqrt{2} gk) ]
+ \frac{2\pi e^{-k}}{1-e^{-k}} P(k;\{x_h\}) - (2\ln 2) \, 2\pi \delta (k)-\nonumber \\
&-& \frac{g^2 \pi k}{\sinh \frac{k}{2}} \int _{0}^{+\infty} dt
 \mathcal{\hat K} ( \sqrt{2} gk, \sqrt{2} gt ) \Bigl [ 2P(t;\{x_h\}) e^{-\frac{t}{2}}+4 e^{-\frac{t}{2}}+ \frac{e^{-\frac{t}{2}}}{\pi}\hat \sigma (t)|_{holes} \Bigr ] \, , \, k>0 \, . \label {tilsigeq}
\ea
Further, we may separate in it the r\^ole of all internal holes (AIH) from the rest (NIH) and, thus, solve by
\be
\hat \sigma (k)|_{holes}= \hat \sigma (k)|_{AIH} + \hat \sigma (k)|_{NIH} \, , \label {second-split}
\ee
where
\ba
\hat \sigma (k)|_{AIH}&=& \frac{2\pi e^{-k}}{1-e^{-k}} \sum _{h=1}^{L-2} \left (\cos k x_h -1 \right ) - \nonumber \\
&-& \frac{g^2 \pi k}{\sinh \frac{k}{2}} \int _{0}^{+\infty} dt
 \mathcal{\hat K} ( \sqrt{2} gk, \sqrt{2} gt ) \Bigl [ 2 e^{-\frac{t}{2}} \sum _{h=1}^{L-2} \left (\cos t x_h -1 \right )+ \frac{e^{-\frac{t}{2}}}{\pi}\hat \sigma (t)|_{AIH} \Bigr ] \, , \quad k>0 \, , \label {tilSigeqA}
\ea
depends on the dynamics of all the internal holes\footnote{It stands to reason that this dynamics is, in its turn, influenced by the interaction with the positions of the roots and the external holes.}, while
\ba
\hat \sigma (k)|_{NIH}&=& \frac{4\pi e^{-k}}{1-e^{-k}}\cos k x_L - \frac{2\pi}{\sinh \frac{k}{2}}J_0(\sqrt{2} gk) -(2\ln 2 ) \, 2\pi \delta (k) -  \nonumber \\
&-& \frac{g^2 \pi k}{\sinh \frac{k}{2}} \int _{0}^{+\infty} dt
 \mathcal{\hat K} ( \sqrt{2} gk, \sqrt{2} gt ) \Bigl [ 4 e^{-\frac{t}{2}}\cos tx_L+ \frac{e^{-\frac{t}{2}}}{\pi}\hat \sigma (t)|_{NIH} \Bigr ] \, , \quad k>0 \, , \label {tilSigeqB0}
\ea
depends on the two external holes positions\footnote{Of course the latter arise also as a consequence of their interaction with the roots and the internal holes, but they can be determined, effectively for the present scope, at leading order, {\it cf.} below.} and a known inhomogeneous term. It is easy to prove that the integral involving $4 e^{-\frac{t}{2}}\cos tx_L$ is of negligible order $O(1/s^2)$ and therefore the final equation to solve for $\hat \sigma (k)|_{NIH}$ is
\ba
\hat \sigma (k)|_{NIH}&=& \frac{4\pi e^{-k}}{1-e^{-k}}\cos k x_L - \frac{2\pi}{\sinh \frac{k}{2}}J_0(\sqrt{2} gk) -(2\ln 2 ) \, 2\pi \delta (k) -  \nonumber \\
&-& \frac{g^2 \pi k}{\sinh \frac{k}{2}} \int _{0}^{+\infty} dt
 \mathcal{\hat K} ( \sqrt{2} gk, \sqrt{2} gt )  \frac{e^{-\frac{t}{2}}}{\pi}\hat \sigma (t)|_{NIH}  \, , \quad k>0 \,  \label {tilSigeqB}.
\ea
We remark the physical nature of this splitting
\be
\hat \sigma (k)= (L-2) \hat \sigma ^{(1)}(k) +  \hat \sigma (k)|_{AIH} + \hat \sigma (k)|_{NIH}  \, ,
\ee
as it clearly identifies the scattering (matrix) of the fundamental excitations, {\it i.e.} the internal holes, in within $\sigma (u)|_{AIH}$. Instead, the rest does not depend on the scattering, but only on the (derivative of the) momentum dispersion relation, $p_h(u)$, of a hole. This is a usual physical interpretation in the theory of the NLIE when neglecting the non-linear terms (infinite size limit) and considering the holes as 'spinon' excitations instead of the magnon excitations \cite{FFGR, FR-XYZ}, which fill in the Fermi sea. In the string theory language, this corresponds to considering excitations over the GKP vacuum \cite{GKP}, instead of those over the BMN vacuum \cite{BMN}. Albeit the exact procedure is so far fully general, we will from now on confine our attention on the non-perturbative $O(6)$ regime, which gives particular account of the above interpretation as it is a relativistic case.

\medskip
Let us first find the solution for $\hat \sigma (k)|_{AIH}$. It is convenient to define the function
\be
S(k)|_{AIH}= \frac{\sinh \frac{k}{2}}{\pi k} \left [ \hat \sigma (k)|_{AIH} - \frac {2\pi e^{-k}}{1-e^{-k}}  \sum _{h=1}^{L-2} \left[ \cos kx_h- 1 \right] \right ] \, , \label {S-sigma}
\ee
which satisfies the integral equation
\be
S(k)|_{AIH}=-g^2 \int _{0}^{+\infty} dt \mathcal{\hat K} ( \sqrt{2} gk, \sqrt{2} gt ) \Bigl [ \frac{\sum \limits _{h=1}^{L-2} \left( \cos tx_h- 1 \right) }{\sinh \frac{t}{2}} + \frac{te^{-\frac{t}{2}}}{\sinh \frac{t}{2}}S(t)|_{AIH} \Bigr ] \, , \quad k > 0 \, . \label {AIH-eq}
\ee
We introduce the Neumann modes $S_p^{\prime}(g)$,
\be
S(k)|_{AIH}=\sum _{p=1}^{\infty}S_p^{\prime}(g) \frac{J_p(\sqrt{2} gk)}{k} \, ,
\label {neumann}
\ee
which, as a consequence of (\ref {AIH-eq}) satisfy the system ($Z_{n,m}(g)$ are defined in (\ref {cZ})):
\ba
S_{2p}^{\prime}(g)&=&-2p \int _{0}^{\infty} \frac{dt}{t}\frac{J_{2p}(\sqrt{2} gt)}{\sinh \frac{t}{2}}
\left [\sum _{h=1}^{L-2} \left (\cos t x_h -1 \right )\right ]  -4p \sum _{m=1}^{\infty}(-1)^{m}Z_{2p,m}(g) S_m^{\prime }(g) \, , \nonumber \\
&& \label {rel3} \\
S_{2p-1}^{\prime }(g)&=&-(2p-1) \int _{0}^{\infty} \frac{dt}{t}\frac{J_{2p-1}(\sqrt{2} gt)}{\sinh \frac{t}{2}}
\left [\sum _{h=1}^{L-2} \left (\cos t x_h -1 \right ) \right ]  -2(2p-1) \sum _{m=1}^{\infty} Z_{2p-1,m}(g) S_m^{\prime}(g) \, . \nonumber
\ea
Concentrating on the first terms in the right hand side of (\ref {rel3}), we develop the cosine functions in power series and exchange the series with the integration. We end up with
\be
S_{p}^{\prime}(g)=-2\pi \sum _{h=1}^{L-2} \sum _{n=1}^{\infty} \frac{(-1)^n (x_h)^{2n}}{(2n)!}
\tilde S_p ^{(n)}(g) \, , \label {S-prime}
\ee
where $\tilde S_p ^{(n)}(g)$ solve system (4.24) of \cite {FGR3}, i.e.
\ba
\tilde S_{2p}^{(n)}(g)&=&2p \int _{0}^{\infty} \frac{dt}{2\pi} t^{2n-1}\frac{J_{2p}(\sqrt{2} gt)}{\sinh \frac{t}{2}}
-4p \sum _{m=1}^{\infty}(-1)^{m}Z_{2p,m}(g) \tilde S_m^{(n)}(g) \, , \nonumber \\
&& \label {rel-fn} \\
\tilde S_{2p-1}^{(n)}(g)&=&(2p-1) \int _{0}^{\infty} \frac{dt}{2\pi} t^{2n-1}\frac{J_{2p-1}(\sqrt{2} gt)}{\sinh \frac{t}{2}}
-2(2p-1) \sum _{m=1}^{\infty} Z_{2p-1,m}(g) \tilde S_m^{(n)}(g) \, . \nonumber
\ea
We now perform the strong coupling limit ($g \rightarrow \infty$) by using the asymptotic expansion of  $\tilde S_p ^{(n)}(g)$ given in (5.3-5.6) of \cite {FGR3}. We can expand as
\be
\frac{\pi k}{\sinh \frac{k}{2}} \sum _{p=1}^{\infty}\tilde S_p ^{(n)}(g) \frac{J_p(\sqrt{2}gk)}{k}= k^{2n}
\frac{e^{\frac{k}{2}}}{2\sinh \frac{k}{2}\cosh k} +\ldots \, , \label {S-tilde-sum}
\ee
where the dots stand for non-perturbative (non-analytic)\footnote{Both terms are referred to any expansion in powers of $1/g$ in a neighbourhood of $g=\infty$.} subleading contributions (that we can neglect here), exponentially small  $\sim e^{-\frac{\sqrt{\lambda}}{4}}$, or more precisely proportional to the mass gap of the embedded O(6) NLSM \cite {AM, FGR1, BK, FGR3}:
\be
m(g)= \frac{2^{\frac{1}{4}}}{\Gamma \left (\frac{5}{4}\right )} \lambda ^{\frac{1}{8}} e^{-\frac{\sqrt{\lambda}}{4}} \left [1+O\left (\frac{1}{\sqrt{\lambda}} \right) \right ] \, .
\label {mass-gap}
\ee
Plugging (\ref {S-tilde-sum}) and (\ref {S-prime}) into (\ref {neumann}), we can re-sum to the cosine function, ending up with the relation:
\be
\frac{\pi k}{\sinh \frac{k}{2}}  \left .
S(k) \right |_{\textrm{AIH}}=\frac{\pi k}{\sinh \frac{k}{2}} \sum _{p=1}^{\infty}S_p^{\prime }(g) \frac{J_p(\sqrt{2} gk)}{k} =-2\pi
\left [\sum _{h=1}^{L-2} \left (\cos k x_h -1 \right ) \right ] \frac{e^{\frac{k}{2}}}{2\sinh \frac{k}{2}\cosh k } + O(m(g)^2)\, . \label {reconstr}
\ee
Putting (\ref {reconstr}) into (\ref {S-sigma}) we end up with the strong coupling expression of $\hat \sigma (k)|_{AIH}$:
\be
\hat \sigma (k)|_{AIH}=2\pi \left [ \frac{e^{-k}}{1-e^{-k}}- \frac{e^{\frac{k}{2}}}{2\sinh \frac{k}{2}\cosh k} \right ]
\sum _{h=1}^{L-2} [\cos k x_{h} -1 ] + O(m(g)^2) \label {fin-AIH}.
\ee
We finally remark that the function
\be
2\pi \left [ \frac{e^{-k}}{1-e^{-k}}- \frac{e^{\frac{k}{2}}}{2\sinh \frac{k}{2}\cosh k} \right ]=
\hat \sigma ^{(1)}_{lim}(k)
\ee
coincides with the strong coupling limit $\hat \sigma ^{(1)}_{lim}(k)$ of the Fourier transform $\hat \sigma ^{(1)}(k)$ of the density corresponding to the first generalised scaling function and also of the $O(6)$ NLSM kernel, i.e. the {\it logarithmic derivative of the scattering amplitude} (cf. below).

\medskip

Passing now to the solution $\hat \sigma (k)|_{NIH}$ to (\ref {tilSigeqB}), we observe how this equation is similar to that for $\gamma (\sqrt{2} gk)$ in \cite {BK,BKnp}, once specialised to the case $L=2$. In this special case the equation for $\gamma (\sqrt{2} gk)$ simplifies very much as it is closed, {\it i.e.} it does not involve another function $\gamma_h$, as it is for our (\ref {tilSigeqB}). Precisely, the term in the second line of (\ref {tilSigeqB})
\be
\hat I(k)|_{NIH}= \hat \sigma (k)|_{NIH}-\frac{4\pi e^{-k}}{1-e^{-k}}\cos k x_L + \frac{2\pi}{\sinh \frac{k}{2}}J_0(\sqrt{2} gk) + (2\ln 2 ) \, 2\pi \delta (k)  \label {Ik},
\ee
can be easily related to the function $\gamma (\sqrt{2} gk)$ used in \cite {BK,BKnp} at $L=2$:
\be
\hat I(k)=2\pi \sqrt{2}\pi g \frac{\gamma (\sqrt{2} gk)|_{L=2}}{\sinh \frac{k}{2}} \, . \label {I-gamma}
\ee
This allows us to adapt the steps of \cite {BK,BKnp} to the evaluation of the strong coupling limit of
$I(u)$ (and, consequently, of $\sigma (u)|_{NIH}$).
We eventually obtain
\be
\sigma (u)|_{NIH}=-\pi m (g) \frac{R(s,g)}{2} \cosh \frac{\pi u}{2} -4 \int _{0}^{+\infty} dt \cos tu \frac{\sinh \frac{t}{2}}{\cosh t} J_{0}(\sqrt{2} gt)+
O(m(g)^3)
\, , \quad |u|< \sqrt{2}g \, , \label {nih1}
\ee
where the 'length' $R(s,g)$ is defined so that
\ba
m (g) R(s,g) &\equiv& \frac{16 \sqrt{2}}{\pi ^2}e^{-\frac{\pi g}{\sqrt{2}}} (\ln 4s + \gamma _E) +
\frac{16 e^{-\frac{\pi g}{\sqrt{2}}}}{\pi}\int _{0}^{+\infty} dt \left [ \frac{J_0 (\sqrt{2} gt)}{e^t-1}
\textrm{Re} \left ( \frac{ie^{it\sqrt{2}g-i\frac{\pi}{4}}}{t+\frac{i\pi}{2}} \right ) - \frac{\sqrt{2}}{\pi} \frac{e^{\frac{t}{2}}}{e^t-1} \right ] - \nonumber \\
&-& \frac{8 g \sqrt{2} e^{-\frac{\pi g}{\sqrt{2}}}}{\pi}\int _{0}^{+\infty} dt \textrm{Re}
\left [ \frac{e^{it-i\frac{\pi}{4}}}{t+i\pi \frac{g}{\sqrt{2}}} \left (\Gamma _+ (t) +i\Gamma _-(t) \right ) \right ]
\ea
and the functions $\Gamma _{\pm}(t)$ are defined by
\be
\Gamma _{\pm}(t)=\gamma _{\pm}(t)\mp \gamma _{\mp}(t) \, \coth \frac{t}{2\sqrt{2}g} \,   ,
\ee
with $\gamma _-(t)=2\sum \limits _{n=1}^{\infty} (2n-1) \gamma _{2n-1}J_{2n-1}(t)$,
$\gamma _+(t)=2\sum \limits _{n=1}^{\infty} 2n \gamma _{2n}J_{2n}(t)$ respectively the 'odd' and 'even' part of the function appearing in (\ref {I-gamma})
\be
\gamma (t)|_{L=2}=\sum _{n=1}^{\infty} 2n \gamma _{n}J_{n}(t) \, .
\ee
The second term in the right hand side of (\ref {nih1}) is easily estimated at strong coupling as
\be
-4 \int _{0}^{+\infty} dt \cos tu \frac{\sinh \frac{t}{2}}{\cosh t} J_{0}(\sqrt{2} gt)\sim
\frac{e^{-\frac{\pi g}{\sqrt{2}}}}{\sqrt{g}} \cdot \cosh \frac{\pi }{2} u   \, .
\ee
For what concerns $R(s,g)$, this quantity has been computed recently in \cite {FRE}
by extending the procedure of \cite{BK, BKnp} (which contemplate general $L>2$ and the presence of $\gamma_h$) to (an equivalent formulation of) the linear integral equation derived in \cite{FIR} which encompasses the sub-leading term $(\ln s)^0$. Then, at the first orders we can write down
\be
R(s,g)= 2 \left ( \ln \mathcal{\bar S} + O(g^{-\frac{3}{4}})\right ) \label {driv} \, ,
\ee
where the parameter $\mathcal{\bar S}$ reads
\be
\mathcal{\bar S} = \frac{2\sqrt{2}s}{g} \, . \label {calbarS}
\ee
Finally, putting all together, we obtain for the driving term in this regime
\be
\sigma (u)|_{NIH}=-\pi m (g) \frac{R(s,g)}{2} \cosh \frac{\pi}{2}u +  O(m(g)^3)
\, , \quad |u|< \sqrt{2}g \, . \label  {fin-NIH}
\ee

\medskip

Upon summing up the expressions (\ref {fin-AIH}, \ref {fin-NIH}) in (\ref {second-split}) and then the latter in (\ref {first-split}) with the cancelation of $(L-2) \hat \sigma ^{(1)}(k)=(L-2) \hat \sigma ^{(1)}_{lim}(k)+O(m(g)^2)$, we obtain, in the $O(6)$ regime, the final equation (in coordinate space) in terms of the holes
\be
\sigma (u)=- \pi m(g) \frac{R(s,g)}{2} \cosh \frac{\pi}{2}u  +\sum _{h=1}^{L-2}
\sigma ^{(1)}_{lim} (u-x_h) + O\left ( \frac{1}{s^2} \right ) + O(m(g)^2) \, , \quad |u|< \sqrt{2}g
\, . \label {fin-sigma}
\ee
In fact, we may interpret this as an equation for $\sigma (u)$ (cf. below for a precise description) since the hole positions (uniquely stemming from (\ref {fin-AIH})) are indeed fixed by $\sigma (u)$. Moreover, we find here an exemplification of the general interpretation (cf. above) of the known term as being the derivative of the momentum, while the excitation terms, coming from $\sigma (u)|_{AIH}$, as being proportional to the derivative of the logarithm of the (relativistic) scattering matrix: $d\ln S(u,x_h)/d u= d\ln S(u-x_h)/d u$. The latter may be identified with an isotopic scattering channel of the $O(6)$ NLSM \cite{Zam,HN}.

Moreover, it is important to remark that the coefficient (\ref {driv}) plays the role of \, 'dynamical length' in the O(6) NLSM description of the subsequent Sections 5 and 6. It contains further quantum fluctuations $O(g^{-\frac{3}{4}})$, which will not affect our results concerning the anomalous dimension
at the order $\frac{1}{\ln \mathcal{\bar S}}$ : indeed, these results are not even sensitive of the $O(g^0)$ corrections, but depend crucially on the peculiar string form $R=2 \ln \mathcal{S}+ O(g^0)=2 \ln (s/g)+O(g^0)$ to all string loops\footnote{For practical reasons we will limit our checks and predictions up to two loops in this work.}, as reported in Section 6.

\medskip

Summarising, we have fixed the equation for the density (\ref{fin-sigma}). We then move on, in next Section, to analyse how it can furnish the anomalous dimension in the present $O(6)$ regime. Crucially, we must observe how the presence of internal holes ($L>2$) is fundamental to render our final equation (\ref{fin-sigma}) non-trivial. In other words, we are clearly analysing  the case $j\neq 0$ (or $\mathcal{J}\neq 0$ for the string), which will lead us in the following to an $O(6)$ NLSM in a magnetic field (or chemical potential) $h\neq 0$. On the contrary, the case $j=0$ (twist $L=2$), {\it i.e.} $\mathcal{J}= 0$ for the string, is deeply different as the $O(6)$ NLSM, without magnetic field, should originate uniquely from the other density portion $\sigma(u)|_{NIH}$ constrained by (\ref {tilSigeqB}), while trivially $\sigma(u)|_{AIH}\equiv 0$. In this context, although in the massive regime $m(g) R \gg 1$ \footnote{In the following we will be more concentrated on the massless (or UV) regime for comparing with semiclassical string expansion ($g\rightarrow +\infty \Rightarrow m \rightarrow 0$).}, the L\"uscher-like term $\sim e^{-m R \cosh \frac{\pi}{2} u}$ has appeared already in \cite {BB} (compare with the derivative of the momentum \label {fin-NIH}) and gives the same leading behaviour of the 'length' $R(s,g)$ (\ref {driv}), delivering us an indirect confirmation of the 'same' $O(6)$ NLSM as string theory low energy action.

In conclusion, all which follows is valid only for higher twist operators, $j\neq 0$, or $\mathcal{J}\neq 0$ for the string.

\section{High spin/strong coupling: the anomalous dimension}
\setcounter{equation}{0}

In order to work out the expression of the anomalous dimension, we shall go back to equation (\ref {Seq}) for the function $S(k)$, as we learned (\ref {S-gamma}), {\it i.e.} $\gamma =2 S(0)$ . We evaluate the nonlinear term by using (\ref {non-lin}) and then set the following definitions,
\be
S(k)=(L-2)S^{(1)}(k)+S(k)|_{NIH}+S(k)|_{AIH} \, ,
\label{S1}
\ee
where $S^{(1)}(k)$ corresponds to the first generalised scaling function ({\it i.e.} $f^{(1)}(g)=2S^{(1)}(0)$ as in \cite{FGR1}),
$S(k)|_{AIH}$ satisfies equation (\ref {AIH-eq}) and $S(k)|_{NIH}$ is the solution of
\ba
S(k)|_{NIH}&=&\frac{2}{k}[1-J_0(\sqrt{2}gk) ]+2g^2 \ln 2  \mathcal{\hat K} ( \sqrt{2} gk, 0)- \\
&-& g^2 \int _{0}^{+\infty} dt \mathcal{\hat K} ( \sqrt{2} gk, \sqrt{2} gt ) \frac{2\cos tx_L -2 e^{-\frac{t}{2}}+te^{-\frac{t}{2}} S(t)|_{NIH}}{\sinh \frac{t}{2}}   \, , \quad k>0 \,\, .
\ea
The function $S(k)|_{NIH}$ has been already studied in \cite {FZ} and it is easily evaluated in $k=0$. Now we concentrate our attention on the Neumann expansion (\ref {neumann}), $S(k)|_{AIH}=\sum \limits _{p=1}^{\infty}S_p^{\prime }(g) \frac{J_p(\sqrt{2} gk)}{k}$, whose contribution in $k=0$ may come only from the $p=1$ mode, $S_1^{\prime }(g)$ (since $J_p(\sqrt{2} gk)\sim k^p$). The latter, purely non-analytic (in $g$), can be found inserting equation (5.1) of \cite {FGR3}, {\it i.e.}
\be
\tilde S_1^{(n)}(g)=\frac{(-1)^{n+1}}{4\pi}\left (\frac{\pi}{2}\right )^{2n} \frac{2 m(g)}{\sqrt{2}g} + O(m(g)^3)\, ,
\ee
into (\ref {S-prime}): thus we easily discover
\be
S_1^{\prime }(g)= \frac{m(g)}{\sqrt{2}g}\sum _{h=1}^{L-2} \left (\cosh \frac{\pi}{2}x_h -1 \right ) + O(m(g)^4)\, .
\ee
Putting this information into (\ref {S1}), we obtain
\be
\gamma = \left . \gamma \right |_{\textrm{NIH}} + (L-2)f^{(1)}(g)
+ m (g) \sum _{h=1}^{L-2} \left (\cosh \frac{\pi}{2}x_h -1 \right )+ O(m(g)^4) \, ,
\ee
where $\gamma |_{\textrm{NIH}}=2\left. S(0)\right |_{NIH} $ is easily written in the notations of the second of \cite {FZ}:
\be
\gamma |_{\textrm{NIH}}= f(g) \ln s + f_{sl}(g)  \, .
\ee
The cusp anomalous dimension $f(g)$ and the 'virtual scaling function' $f_{sl}(g)$ were found to be governed at any coupling by linear integral equations respectively in \cite{BES} and \cite{BFR}, whose strong coupling expansions were developed in \cite {BKK} and \cite {FZ}. Besides, we find useful the strong coupling exact behaviour of \cite {FGR1,BK}
\be
f^{(1)}(g)=-1+m(g)+ O(m(g)^2)\, ,
\ee
in that it allows us to simplify, in this regime,  the expression for
\be
\gamma =  f(g) \ln s - (L-2)  + f_{sl}(g)
+ m (g) \sum _{h=1}^{L-2} \cosh \frac{\pi}{2}x_h + O(m(g)^3)\, . \label {andime}
\ee
Equations (\ref {fin-sigma}, \ref {andime}) can be more easily handled if we express them in an integral form. This will be performed in next section.

\section{Non-Linear Integral equation in the $O(6)$ regime}
\setcounter{equation}{0}

The different sums over the internal holes in equations (\ref {fin-sigma}, \ref {andime}) may be converted into an integral form by using the residue theorem \cite{FMQR}, as summarised by (\ref {cauchy}) in Section 2.
Applying (\ref {cauchy}) to the density of holes and discarding the undesired $O(1/s^2)$, $O(m(g)^2)$ terms, we obtain the following equation
\ba
\sigma (u)&=&- \pi m(g) \frac{R(s,g)}{2} \cosh \frac{\pi}{2}u  - \int _{-c}^c \frac{dv}{2\pi} \sigma ^{(1)}_{lim}(u-v)
\sigma (v)+ \nonumber \\
&+& \int _{-c}^c \frac{dv}{\pi}\sigma ^{(1)}_{lim}(u-v) \ \frac{d}{dv}{\mbox {Im}} \ln
[1+(-1)^{L} \, e^{iZ(v-i0^+)} ] \, , \quad |u|< \sqrt{2}g \, . \label {nlie-den}
\ea
On the other hand, for the anomalous dimension we have the relation
\ba
\gamma &=&  f(g) \ln s - (L-2)  + f_{sl}(g) - m(g) \int _{-c}^c \frac{dv}{2\pi}  \cosh \frac{\pi}{2}v
\ \sigma (v) + \nonumber \\
&+& m(g) \int _{-c}^c \frac{dv}{\pi} \cosh \frac{\pi}{2}v  \ \frac{d}{dv}{\mbox {Im}} \ln
[1+(-1)^{L} \, e^{iZ(v-i0^+)}] \, . \label {nlie-andim}
\ea
The nonlinear terms in the second lines of (\ref {nlie-den}, \ref {nlie-andim}) highlight the fact that
these expressions assume the form of a non-linear perturbation of the $O(6)$ NLSM in the thermodynamic limit (infinite size).

In the two relations above everything depends relativistically on the two generic quantities 'mass' $m$ and size $R$. With respect to the AdS/CFT variables, while $m(g)$ gets exponentially small as $g\rightarrow +\infty$, the size $R(s,g)=2 \ln (s/g) + ...$ depends on the ratio between the two large quantities $s$ and $g$. The latter is assumed to be very large in the following expansions without any assumption on its value with respect to $m$. More precisely, we will expand the non-linear integrals in (\ref {nlie-den}, \ref {nlie-andim}) as in \cite{FIR}, in order to obtain equations depending only on the density of holes $\sigma (u)$. With this aim in mind, we reckon convenient to introduce the quantities
\ba
&& \chi (\theta )= - \frac{2}{\pi} \sigma (u) \, , \quad \theta = \frac{\pi}{2}u \, , \quad B = \frac{\pi}{2} c \, , \quad K(\theta )=- \frac{1}{\pi ^2} \sigma ^{(-1,1)}_{lim}(u) \, ,  \nonumber \\
&& K(\theta)= \frac{1}{4\pi ^2} \left [ \psi \left ( 1-\frac{i\theta}{2\pi} \right )+
\psi \left ( 1+\frac{i\theta}{2\pi} \right )- \psi \left ( \frac{1}{2}-\frac{i\theta}{2\pi} \right )-
\psi \left ( \frac{1}{2}+\frac{i\theta}{2\pi} \right ) + \frac{2\pi}{\cosh \theta} \right ] \, ,
\label {newquant}
\ea
which make contact with $O(6)$ NLSM conventions ($\chi(\theta)$, for instance, is the density of excitations). In terms of (\ref {newquant}), the integral expressions (\ref {nlie-den}, \ref {nlie-andim}) take this expanded form, respectively
\ba
\chi (\theta)&=& m(g) \ R(s,g) \cosh \theta +
\int _{-B}^{B}
d\theta ' K(\theta - \theta ') \chi (\theta ') +   \label {chi2} \\
&+& \pi \sum _{k=0}^{\infty} \frac{(2\pi )^{2k+1}}{(2k+2)!} B_{2k+2}\left (\frac{1}{2}\right ) \left [ \left ( \frac{1}{\chi (\theta ')} \frac{d}{d\theta '} \right )^{2k+1} [ K(\theta + \theta ')+K(\theta - \theta ') ] \right ]^{\theta '=B}_{\theta '=-B} \, ,  \quad |\theta|< \frac{\pi g}{\sqrt{2}} \, ,
\nonumber
\ea
\be
\varepsilon = m(g) \int _{-B}^{B} \frac{d\theta }{2\pi} \chi (\theta) \cosh \theta +
m(g) \sum _{k=0}^{\infty} \frac{(2\pi )^{2k+1}}{(2k+2)!} B_{2k+2}\left (\frac{1}{2}\right ) \left [ \left ( \frac{1}{\chi (\theta)} \frac{d}{d\theta} \right )^{2k+1} \cosh \theta \right ]^{\theta = B}_{\theta = -B}
\label {epsilon2} \, ,
\ee
where
\be
\varepsilon = \gamma - f(g) \ln s  + (L-2) - f_{sl}(g) \, \label {vareps-def}
\ee
and $B_k(x)$ is the Bernoulli polynomial.
For these expressions can be interpreted as perturbations at all (power-like) orders \footnote{In this sense, we mean here the non-linearity. Also, the expanded integrals are non-linear as in \cite{FIR}.} of the small parameter  $\frac{1}{R(s,g)}= \frac{1}{2 \ln \mathcal {\bar S}}+...$\footnote{This model does not coincide with the finite size $O(6)$ NLSM, but represents for sure the starting point to understand if and how the latter appears in AdS/CFT (for finite length).}. Equations (\ref {chi2}, \ref {epsilon2}) are supplemented by the condition
\be
L-2=\int _{-B}^{B} \frac{d\theta}{2\pi} \chi (\theta ) \, , \quad  \label {j2}
\ee
which counts the internal holes. Introducing the string 'renormalised' parameter (cf. with its string perturbation analogue (\ref{ell-string}))
\be
\ell= \frac{2\pi (L-2) }{ \sqrt{\lambda} R(s,g) }\, ,
\label {ell-par}
\ee
we write (\ref {j2}) in the alternative form:
\be
\frac{\sqrt{\lambda } \ell R(s,g) }{2\pi}=\int _{-B}^{B} \frac{d\theta}{2\pi} \chi (\theta ) \, . \quad  \label {j2-1}
\ee
The structure of (\ref {chi2}, \ref {epsilon2}, \ref {j2-1}) implies that the anomalous dimension organises in powers of $R$. The leading order is proportional to the size $R \simeq 2 \ln s$ and is commonly called 'bulk' term as it is the only one which survives in the thermodynamic (infinite size) limit (after dividing by $R$): the $R^0$ term is absent in closed (periodic) systems as it is usually associated to the presence of boundaries, and the rest vanishes for $R\rightarrow \infty$. Interestingly, only odd powers of $R$ (in specific $R$, $\frac{1}{R}$, $\frac{1}{R^3}$, ...) appear in the large $R$ expansion of $\chi (\theta)$ and $\varepsilon$
\footnote {Similarly, only even powers of $R$ ({\it i.e.} $R^0$, $\frac{1}{R^2}$, ...) appear in the large $R$ expansion of $B$.}.

As mentioned since the first section, we have obtained our basic equations in the so-called $O(6)$ NLSM regime $j\ll \sqrt{\lambda}$. Within this domain, we can have two possible regimes of $j$ with respect to $m(g)$. 1) The first one, $j\ll m$, has been used to derive and analyse the infinite volume $O(6)$ previously in \cite{FGR1, FGR3}: following this, we may simply generalise the investigation to the present case for finite size. This is also the strong coupling regime of the $O(6)$ quantum field theory. 2) The second one, $m\ll j$, is instead the UV (or massless) expansion which also corresponds to the weak coupling regime for the field theory. Analogously, it is this expansion which allows us comparison with the $\mathcal{J}\neq 0$ string perturbation series. Therefore, as for the bulk energy
we re-discover the $O(6)$ NLSM solution by Alday and Maldacena for the smallest order $\ell^2$  \cite {AM} in a different way if compared with \cite{FGR1,BK,FGR3}. Hence, we can use the results by \cite {HN} on NLSMs (obtained by the scattering Bethe Ansatz in the thermodynamic limit) and write down the energy as
\be
\varepsilon (\lambda, \ell)= R(s,g) \ \ell ^2 \left [ \frac{\sqrt{\lambda}}{4\pi } - \frac{1}{\pi} \ln \ell + \frac{3}{4\pi}+ O\left ( \frac{1}{\sqrt{\lambda}} \right ) \right ] \label {vareps-1} \, ,
\ee
which agrees with the one-loop string computation \cite {FTT}\footnote {It is important to remark that the one loop result of \cite {FTT} was exactly confirmed by manipulating the ABA equations in matrix-model-like methodology \cite {CK}.}.

The order $R^0$ would be a boundary term and is absent, as reasonable for a closed system. This result agrees with classical energy given by (2.13) of \cite {BFTT}, which show that the first nonzero contribution is proportional to $R^0 \cdot \ell ^4$ and, consequently, is beyond the $O(6)$ low energy limit.

On the contrary, the next order, {\it i.e.} the $1/R$ finite size correction or Casimir effect, will be demonstrated to be very intriguing in the next section.

\section{Finite size effect: $1/R\sim1/\ln \mathcal{\bar S}$ order and $O(6)$ NLSM}
\setcounter{equation}{0}

To compute this term at high spin, we are allowed to truncate (\ref {chi2}, \ref {epsilon2}, \ref {j2-1}) at the order $1/R(s,g)\sim 1/ \ln \mathcal {\bar S}$, giving rise to the equations:
\ba
g (\theta)- \int _{-B}^{B}d\theta ' K(\theta -\theta ') g(\theta ') = m \cosh \theta
- \frac{ \pi ^2}{6 g(B) R^2} \frac{d}{dB}[K(B-\theta)+K(B+\theta)]
\label {gi3} \, ,
\ea
\be
E(\rho)  = m (g)\int _{-B}^{B} \frac{d\theta }{2\pi} \cosh \theta \ g(\theta) -
\frac{m \pi}{6} \frac{\sinh B}{ g(B) R^2}
\label {En3} \, ,
\ee
\be
\rho =\int _{-B}^{B} \frac{d\theta}{2\pi} g (\theta )   \label {rho3} \, ,
\ee
with the identifications
\be
\quad E (\rho)= \frac{\varepsilon }{R} \, , \quad g(\theta)= \frac{\chi (\theta )}{R}\, , \quad \rho = \frac{L-2}{ R}  \,  . \label {identi2}
\ee
In this manner we realise that at this order -- but at the next orders as well -- the theory may be interpreted as a (small) perturbation of the usual $O(6)$ NLSM, provided the perturbation parameter $1/(R(s,g))$ is small; in this specific case we only consider the 'linear' perturbation, albeit we well could go further with the quadratic terms. Therefore we adapt a standard procedure for two-dimensional relativistic (integrable) scattering models in the thermodynamic limit (a concise summary of the relevant passages is presented in Appendix \ref {useful}, according to \cite{HN} and preceding references therein) and consider the free energy, {i.e.} the minimum (w.r.t. $\rho$) of the quantity
\be
E(\rho)-h\rho=\int _{-B}^{B} \frac{d\theta }{2\pi} \left (m \cosh \theta -h\right ) g(\theta) -
\frac{m \pi}{6 R^2} \frac{\sinh B}{ g(B)} \, ,
\ee
which amounts to perform the Legendre transform defining the 'external field' $h$ (as function of $\rho$ and vice versa). A standard calculation shows that
\ba
E(\rho)-h\rho &=& -\int _{-B}^{B} \frac{d\theta }{2\pi}
\left (m \cosh \theta - \frac{ \pi ^2}{6 g(B)
R^2} \frac{d}{dB}[K(B-\theta)+K(B+\theta)]\right )
\epsilon (\theta ) - \nonumber \\
&-&  \frac{m \pi}{6 R^2} \frac{\sinh B}{ g(B)} \label {E-rho} \, ,
\ea
where $\epsilon (\theta )$ satisfies the usual equation
\be \epsilon (\theta )- \int
_{-B}^{B}d\theta ' K(\theta -\theta ') \epsilon (\theta
')=h-m \cosh \theta \label {eps2} \, .
\ee
Using (\ref {eps2}), we rewrite (\ref {E-rho}) as follows
\be
E(\rho)-h\rho=-\int
_{-B}^{B} \frac{d\theta }{2\pi} m  \cosh \theta \, \epsilon
(\theta) +\frac{\pi}{6 R^2} \frac{\epsilon '(B)}{ g(B)} \, , \label {E-rho-1}
\ee
where we use the short notation
\be
\epsilon ' (\theta) = \frac{d}{d\theta }\epsilon (\theta) \Rightarrow \epsilon '(B)= \left. \frac{d}{d\theta }\epsilon (\theta) \right |_{\theta =B} \, .
\ee
As anticipated, we define the free energy as the minimum (on $\rho$) 
\be
f(h)={ \textrm{min}}_{\rho}[E(\rho)-h\rho]
\label {fh-1} \, .
\ee
Of course, this implies that $B$ and $h$ are not independent but constrained by the relation $\frac{\partial }{\partial B}[E(\rho(B))-h\rho(B)] =0$. We work on this relation by making explicit the dependence of $\epsilon (B,h;\theta)$ on the parameters $B$ and $h$ and of $g(B;\theta)$ on $B$ and by using
(\ref {E-rho}). We obtain the (perturbed) boundary condition
\be
g(B;B) \epsilon (B,h;B) +
\frac{\pi ^2}{6R^2} \left [ \frac{d}{dB} \frac{m \sinh B}{g(B;B)}- \int _{-B}^{B} \frac{d\theta
}{2} \epsilon (B,h;\theta) \frac{d}{dB} \frac{\frac{d}{dB}[K(B+\theta )+ K(B-\theta)]}{g(B;B)}  \right ]=0 \, . \label {h-rho-0}
 \ee
We now perform the derivatives with respect to $B$ in the square bracket and obtain two types of terms: that given by differentiating $g(B;B)$ is proportional to $\epsilon '(B,h;B)$, the other one to $\epsilon ''(B,h;B)$. We end up with the condition
\be
\epsilon (B,h;B)g(B;B)+ \frac{\pi ^2}{6 R^2} \frac{\epsilon ' (B,h;B)\frac{d}{dB}g(B;B)-g(B;B)\epsilon ''(B,h;B)}{g(B;B)^2} =0 \, \label {h-rho-1} \, ,
\ee
which furnishes the Fermi rapidity $B=B(h)$ in terms of the independent variable $h$, the 'magnetic field' or 'chemical potential'. To this aim, we find it convenient to split the solution as
\be
B(h)=B^{(0)}(h) +B^{(2)}(h) \, ,
\label{B-split}
\ee
where $B^{(0)}(h)$ takes into account the leading O(6) contribution $O(R^0)$ (i.e. it is the usual unperturbed relation), while its first correction $B^{(2)}(h)$ is $O(1/R^2)$. At the leading order $O(R^0)$ equation (\ref {h-rho-1}) reduces to the 'old' $\epsilon (B^{(0)},h;B^{(0)})=0$. This condition fixes $B^{(0)}=B^{(0)}(h)$ as function of $h$ and, in addition, implies also that
\be
\frac{\partial}{\partial B^{(0)}}\epsilon (B^{(0)},h;\theta) \Bigl |_{B^{(0)}=B^{(0)}(h)}=0 \, , \label {cond-1-bis}
\ee
since the function $\frac{\partial}{\partial B^{(0)}}\epsilon (B^{(0)},h;\theta)$ satisfies
the integral equation
\be
\frac{\partial}{\partial B^{(0)}}\epsilon (B^{(0)},h;\theta)=
[K(\theta -B^{(0)})+K(\theta +B^{(0)})]\epsilon (B^{(0)},h;B^{(0)})+ \int _{-B^{(0)}}^{B^{(0)}}d\theta ' K(\theta -\theta ')
\frac{\partial}{\partial B^{(0)}}\epsilon (B^{(0)},h;\theta ') \, .
\ee
The function $B^{(2)}(h)$
is determined by condition (\ref {h-rho-1}) at order $O(1/R^2)$:
\be
B^{(2)}(h)=- \frac{\pi ^2}{6 R^2} \frac{\frac{d}{dB^{(0)}(h)}g(B^{(0)}(h);B^{(0)}(h))- g(B^{(0)}(h);B^{(0)}(h))\frac{\epsilon ''(B^{(0)}(h),h;B^{(0)}(h))}{\epsilon '(B^{(0)}(h),h;B^{(0)}(h))}} {[g(B^{(0)}(h);B^{(0)}(h))]^3}  \, . \label {cond-2}
\ee
Now, using (\ref {E-rho-1}), we express $f(h)$ in terms of $B^{(0)}(h)$ and $B^{(2)}(h)$. We have
the expression
\ba
f(h)&=&-\int _{-B^{(0)}(h)}^{B^{(0)}(h)} \frac{d\theta }{2\pi} m \cosh
\theta  \ \epsilon (B^{(0)}(h),h;\theta) + \frac{\pi}{6 R^2} \frac{\epsilon '(B^{(0)}(h),h;B^{(0)}(h))}{g(B^{(0)}(h);B^{(0)}(h))} - \nonumber \\
&-& \int _{-B^{(0)}(h)}^{B^{(0)}(h)} \frac{d\theta }{2\pi} m \cosh \theta
\left [ \frac{\partial}{\partial B^{(0)}(h) }\epsilon (B^{(0)}(h),h;\theta)
\right ] B^{(2)}(h) - \nonumber \\
&-& \frac{m  \cosh B^{(0)}(h)}{\pi}\epsilon (B^{(0)}(h),h;B^{(0)}(h))
B^{(2)}(h) \, , \nonumber
\ea
in which we must impose conditions $\epsilon (B^{(0)}(h),h;B^{(0)}(h))=0$ and (\ref {cond-1-bis}).
Therefore, the expression of the free energy in terms of the independent variable
$h$ simplifies as
\ba
f(h)&=& -\int _{-B^{(0)}(h)}^{B^{(0)}(h)} \frac{d\theta }{2\pi} m
\cosh \theta \ \epsilon (B^{(0)}(h),h;\theta) +\frac{\pi}{6 R^2}  \frac{\epsilon '(B^{(0)}(h),h;B^{(0)}(h))}{g(B^{(0)}(h);B^{(0)}(h))}  = \nonumber \\
&=& f^{(0)}(h) +\frac{\pi}{6 R^2}  \frac{\epsilon '(B^{(0)}(h),h;B^{(0)}(h))}{g(B^{(0)}(h);B^{(0)}(h))} \, , \label {f-f0}
\ea
which shows explicitly the correction to the infinite size value $f^{(0)}(h)$. This expression holds at all loops and in terms of (infinite size) O(6) NLSM quantities only, in that $B^{(0)}=B^{(0)}(h)$ is the usual unperturbed relation by virtue of (\ref{B-split}).  The $\frac{1}{R^2}$ term, the so-called Casimir free energy, enjoys a peculiar expression as logarithmic derivative of the charge density $\rho (B^{(0)})$ (with respect to $B^{(0)}$); for by means of the convenient relation (\ref {final-rel}) (derived in Appendix \ref {useful}) we may write
\be
f(h)=f^{(0)}(h)-\frac{\pi}{6  R^2}\frac{\rho}{\frac{d \rho}{dB^{(0)}}}=f^{(0)}(h)-\frac{\pi}{6  R^2}\frac{dB^{(0)}}{d (\ln \frac{\rho}{m})}  \, \label {f-f0-1} \,\, ,
\ee
where all the relations, $\rho=\rho(B^{(0)})$, $B^{(0)}=B^{(0)}(h)$, $\rho=\rho(h)$ (different from (\ref{rho-h}): this represent a little abuse of notation \footnote{This is due to the fact that so far we have considered $h$ as an independent variable, thus independent of $R$. Instead in the end, from (\ref{rho-h}) on, we wish to re-express $h=h(\rho)$ in terms of $\rho$, which is the natural  independent AdS/CFT variable. Therefore, $h(\rho)$ becomes corrected at order $O(1/R^2)$ according to (\ref{rho-h}).}), are the unperturbed ones. Thanks to (\ref {f-f0-1}) we can go to the energy $E$ as function of $\rho$ and then the natural AdS/CFT variable $\ell $
\be
E(\ell)=f(h)+\rho h = f^{(0)} (h)+\rho h - \frac{\pi}{6 R^2}\frac{\rho}{\frac{d\rho}{dB^{(0)}}} (h) \, , \label {E-fin}
\ee
by means of the (perturbed) Legendre transformation $h=h(\rho)$:
\be
\rho = - \frac{df}{dh}= - \frac{df^{(0)}}{dh}+ \frac{\pi}{6 R^2} \frac{d}{dh} \left( \frac{\rho}{\frac{d\rho}{d B^{(0)}}} (h) \right) \, ,
\label {rho-h}
\ee
and the simple identification 
\be
\rho = \frac{\sqrt{\lambda }}{2\pi} \ell \,. 
\ee
As anticipated in the note, to solve the previous transformation with respect to $h$ we need to assume the $1/R$ expansion $h(\ell)=h^{(0)}(\ell)+h^{(2)}(\ell)$, with $h^{(0)}(\ell)$ the (bulk) $O(6)$ NLSM function and the small perturbation $h^{(2)}(\ell)= O(1/R^2)$\footnote{Notice that the semiclassical string parameter $\ell_s=\ell+\dots$ plays the role of the (infinite size) O(6) NLSM magnetic field $h^{(0)}=\ell+\dots$ \, . Therefore, the limit $\ell_s \rightarrow 0$ causes a sort of double vacuum transition: one vacuum transition is the usual one happening in the $O(6)$ NLSM when the magnetic field $h^{(0)}$ vanishes; the second one corresponds to the above depicted peculiarity of the AdS/CFT ABA, namely the change from the density $\sigma (u)|_{AIH}$ to  $\sigma (u)|_{NIH}$ once the minimal twist is reached.}. Upon plugging this small splitting into (\ref {E-fin}) (and using (\ref {rho-h}) for a cancellation) up to the $1/R^2$ order, this amounts to substituting $h^{(0)}$ for $h$
\be
E(\ell)=E^{(0)}(\ell) - \frac{\pi}{6 R^2}\frac{\rho}{\frac{d\rho}{dB^{(0)}}} (h^{(0)}(\ell)) \, ,
\label {1/R2}
\ee
and thus obtaining the exact correction to the (bulk) energy of the (infinite size) O(6) NLSM, $E^{(0)}(\ell)$ of \cite{AM} at order $1/R^2$. Thus, this Casimir energy\footnote{How to obtain it from the $O(6)$ NLSM TBA equations \cite{ZFBH} represents an intriguing question, as we cannot see any problem to extend the validity of the $O(6)$ NLSM description for the low energy string theory, even at finite size.}
\be
 - \frac{\pi}{6 R^2}\frac{\rho}{\frac{d\rho}{dB^{(0)}}} (h^{(0)}(\ell)) = - \frac{\pi}{6 R^2} \frac{d}{d(\ln \frac{\rho}{m})} B^{(0)}\left ( \frac{\rho}{m} \right ) \, , \label {first-corr}
\ee
holds at all loops and, of course, depends uniquely on $O(6)$ (bulk) NLSM data and more precisely only on the dimensionless ratio $\rho/m$\footnote{The last expression as the derivative of the Fermi rapidity $B^{(0)}(\tau)$ with respect to the renormalisation group 'time' $\tau=\ln (\rho/m)$ is particularly suggestive.}. Incidentally, result (\ref {first-corr}) is confirmed
by independent calculations reported in Appendix \ref {Bexp}, which do not make use of the free energy $f(h)$.

For comparing (\ref {first-corr}) with string theory expansion, we need to expand it at large
$\rho/m$ ({\it i.e.} the ultraviolet (UV)) by simply taking the derivative of the expansion (\ref {Brhom}) for $B^{(0)}\left ( \frac{\rho}{m} \right)$\footnote{It may also be obtained easily from results of \cite {VOL}. Moreover, the latter allow us to go further in loop expansion and show the absence of unwanted $\frac{\ln \lambda}{\lambda }$ terms.}
\be
\frac{d}{d(\ln \frac{\rho}{m})} B^{(0)}\left ( \frac{\rho}{m} \right )=1-\frac{1}{2\ln \frac{\rho}{m}}
+ O \left ( \frac{\ln \ln \frac{\rho}{m}}{\ln ^2 \frac{\rho}{m}} \right ) \, .
\ee
Finally, we need to resume the dependence of the mass $m=m(g)$ on the 't Hooft coupling, (\ref {mass-gap}), and write the Casimir term at one and two string loops as
\be
- \frac{\pi}{6 R^2(s,g)}
\left [1- \frac{2}{\sqrt{\lambda}} +
\ldots  \right ] \, .
\ee
Eventually, going back to the energy density, we conclude that
\ba
\varepsilon (\lambda, \ell)&=& R(s,g) \ \ell ^2 \left [ \frac{\sqrt{\lambda}}{4\pi } - \frac{1}{\pi} \ln \ell + \frac{3}{4\pi}+O \left ( \frac{1}{\lambda ^{1/2}}\right) \right ] - \nonumber \\
&-&  \frac{\pi}{6 R(s,g)} \left [1- \frac{2}{\sqrt{\lambda}}
+ O\left ( \frac{1}{\lambda }\right) \right ] \label {eps-fin}\, .
\ea
This equation states that the ABA contribution to the anomalous dimension at the order $\frac{1}{R}$, {\it i.e.} the Casimir energy takes the value
\be
\left . \gamma \right |_{\frac{1}{R}}= - \frac{\pi}{6 R(s,g)}\left [1- \frac{2}{\sqrt{\lambda}} +
O\left ( \frac{1}{\lambda }\right) \right ]  \nonumber \\
\ee
or consequently
\be
\left . \gamma \right |_{\frac{1}{\ln \mathcal{\bar S}}}= - \frac{\pi}{12 \ln \mathcal{\bar S}}\left
[1- \frac{2}{\sqrt{\lambda}}
+ O\left ( \frac{1}{\lambda }\right) \right ] \, .
\label{gamma-1}
\ee
At this point we shall comment on the previous expression. Since we have derived it from ABA, we can compare it only with the string case
$\mathcal{J}\neq 0$ (with $\mathcal{J}=L/\sqrt{\lambda}$), where the exponential
L\"uscher-like\footnote{Actually, the original L\"uscher exponential correction concerns
the behaviour of the mass-gap (first excited state) at large size.} term $\sim e^{-(const.) \cal{J}}$ does not contribute and takes account of four light modes \cite{GRRT}. Anticipating some of the considerations of next Section, we may observe that the $1$ in square brackets in the previous expansions reveals one massless (scalar) mode which contributes to the Casimir order $1/\ln \mathcal{\bar S}$ according to \cite{Bloete:1986qm}. This amounts to the one-loop contribution $-\pi /12$ in agreement with semiclassical string \cite {BDFPT, GRRT} and algebraic curve \cite {GSSV} computations.
In addition, we predict here $\pi/6$ at two loops as coming from ABA ($\mathcal{J}\neq 0$); this result does not enjoy by now a comparable outcome from either algebraic curve either string expansion. The latter furnishes a (different) hint, as long as $\mathcal{J}=0$, which may be not so firm as it depends on an ambiguous regularisation \cite{GRRT}. The difference of sign between one and two loops could be symptomatic of an alternating sign series,
which is a typical feature of loop expansions in quantum field theories.

Eventually, we may remark how our procedure allows us to go further both in the (string) loop expansion (as we have the exact expression (\ref{first-corr})) and in the $1/R$ perturbation.

\section{Summary and outlook}
\setcounter{equation}{0}

We have started by the nonlinear integral equation for the counting function or its derivative, the density of roots and holes, (\ref{sigmaeq}). It is equivalent to the ABA for twist operators of the $sl(2)$ scalar sector in ${\cal N}=4$ SYM. Furthermore, it is extremely effective to perform the high spin and strong coupling limit in the string scaling regime (\ref{Slimit}), in particular when aiming at the hole NLIE (\ref{sigmaeq2}) upon discarding non-linear terms $\sim 1/\mathcal{S}^2$ ({\it cf.} Section 3). Then, we have specialised our treatment to the non-perturbative small $\ell_s \ll 1$ regime which, besides, gives the lowest order in $\ell_s$ of string perturbation theory, since the the hole NLIE (\ref{sigmaeq2}) encompasses also the string perturbative expansion. The non-perturbative nature is given by the (rest) mass $m\sim e^{-\frac{\sqrt{\lambda}}{4}}$ of the scalar excitations in (\ref{sl2op}), which enjoy relativistic energy and momentum, as long as they are small enough to be comparable with the exponentially small mass $m$. Eventually, we have arrived to equation (\ref {fin-sigma}) for the density (Section 3) and (\ref {andime}) for the anomalous dimension (Section 4), which could be rewritten (Section 5) in an integral form (\ref {chi2}, \ref {epsilon2}), together with the condition (\ref {j2-1}), counting the internal holes. The non linear integral equations (\ref {chi2}, \ref {epsilon2}, \ref {j2-1}) enjoy already a suitable form to be interpreted as an all order perturbation of the $O(6)$ non-linear sigma model by the small size inverse $1/R$ and constitute the starting point for an expansion of the anomalous dimension in inverse powers $R^{-2r-1}$, $r\geq 0$.

\medskip

On the basis of this new 'model', we have obtained summarily the following results. As already stated, at leading order $R$ we have found anew that the bulk term is given by the infinite size $O(6)$ NLSM as predicted by Alday and Maldacena \cite{AM} (in agreement with string theory world-sheet expansion \cite{FTT}). Yet, we have obtained this by a new route which highlights the physical origin of the model (from ABA), namely how the internal holes interact according to a scattering amplitude of the $O(6)$ model \cite{Zam, HN}. Furthermore, as expected for a periodic system, there is no 'boundary' term $R^0$.

Instead, the first finite size term $\frac{\pi}{\textrm{6 (size)}} c = \frac{\pi}{6 R(s,g)} c \cong \frac{\pi c}{12 \ln \mathcal{\bar S}}$ (discussed in Section 6) yields the central charge $c$ of the massless limiting 2D CFT \cite{Bloete:1986qm}. In other words, it counts, at one loop, the number of the massless UV degrees of freedom: in the well-known case of the UV limit
$m\rightarrow 0$ of the 'pure' $O(6)$ NLSM (without magnetic field $h^{(0)}\simeq \ell=0$ (twist two)\footnote{We do not linger here on the obvious problem of the order of the limits $\ell=0$ and $m\rightarrow 0$, which in this way must be taken, as motivated above (cf. also \cite{BB}).}), it amounts to the value $c=5$ as all the five scalars contribute \cite{ZFBH}; on the contrary, in the ABA theory, with a mass scale $\ell\simeq h^{(0)}$, $c=1$ appeared since out of five only the scalar along the magnetic field $h^{(0)}$ is indeed massless. In fact, this latter case corresponds to the string perturbative diagrammatics where the mass of four (scalar) modes $M\sim \ell_s\simeq h^{(0)}$ does not vanish, so that in the large size limit  $M R \sim \ell_s \ln \mathcal{\bar S} \sim \mathcal{J} \gg 1$ these four modes contribute to the exponentially smaller  L\"uscher-like terms $\sim e^{-(const.) \mathcal{J}}$ \cite{GRRT}(which are not captured by ABA). These contributions should be given by the wrapping TBA corrections and yield coincidence with the $\mathcal{J}=0$ case of semiclassical string theory result $-(1+4)\pi/6$ of \cite{GRRT}, thus confirming the fully massless limit of the $O(6)$ NLSM. We are confident that this extra contribution may be dealt with the technique developed in \cite{FIR} and this work as well. Finally, we can produce, at two loops, a prediction which may guide in future string computations and in specific regularisation procedures along the lines of \cite{GRRT}. At any rate, this result ought to be wrapping affected, too, since it differs from the vanishing result of a $\mathcal{J}=0$ string computation \cite{GRRT}. Nevertheless, the latter calculation severely depends on an unclear (at least to us) regularisation which may be a source of ambiguity \cite{GRRT}. Hence, {\it a fortiori} a definite TBA computation would be orientating for string computations and represent a fine test of the AdS/CFT TBA (in the $sl(2)$ sector).

Of course, it is just a matter of computing to go further to higher loops in Section 6 without any further knowledge of the $g$ expansion of $R$. More complicated, but still possible is the analysis of the higher $R$ powers by means of the exact equations (\ref {chi2}, \ref {epsilon2}, \ref {j2-1}). Eventually, we must recall how we produce here only the leading terms in $\ell$, as we are in the $O(6)$ regime, but the hole NLIE (\ref {sigmaeq2}) could be used to go further and compare with string perturbation theory. In particular, we should see the arising of a $R^0$ term. The latter would have the physical interpretation of a boundary term ('vacuum degeneracy') and might enjoy an interpretation and an exact expression as for relativistic models (cf. \cite{CDFT} and references therein).

\medskip

{\bf Acknowledgements}
It is our pleasure to thank J. Balog, B. Basso, A. Belitsky, G. Korchemsky, S. Giombi, S. Piscaglia, R. Roiban and K. Zarembo for illuminating discussions. We also thank L. Freyhult for helpful comments and for sending to one of us (D.F.) a draft version of \cite {FRE}. This work has been partially supported by INFN grants {\it Iniziative specifiche FI11} (D.F.) and {\it PI14}, the ESF Network {\it HoloGrav 'Holographic methods for strongly coupled systems'} (09-RNP-092 (PESC)) (D.F.) and the MIUR- PRIN contract 2009-KHZKRX (D.F.).

\appendix

\section{Exact relations for integrable relativistic models} \label {useful}
\setcounter{equation}{0}

In this appendix we give some useful exact relations which hold in two dimensional integrable relativistic models in the thermodynamic limit. These relations may eventually be specialised to NLSMs and in particular to the $O(6)$ NLSM by simply using expression (\ref {newquant}) for the generic kernel $K(\theta)$. In order to make notations simpler, we will write only the dependence on $\theta$ and we will drop the index $^{(0)}$ we used in Section 6 to denote $O(6)$ NLSM quantities in distinction to their perturbations.

The ground state is obtained by filling in the Fermi see by (interacting) particles of mass $m$ (the mass gap and rapidity $\theta$ from $-B$ to $B$: hence they have 'local' density $g(\theta)$ constrained by the thermodynamic limit of the Bethe Ansatz equations, {\it i.e.} the linear integral equation
\ba
g(\theta)= m \cosh \theta + \int
_{-B}^{B}d\theta ' K(\theta -\theta ') g (\theta ')  \, , \label {gequat}
\ea
constrained ('global' condition) by fixed total density (\cite{HN} and references therein)
\be
\rho = \int _{-B}^{B} \frac{d \theta}{2\pi} g(\theta) \, . \label {rho3'}
\ee
This relation determines the Fermi rapidity $B=B(\rho)$ in terms of a given total density (or number of particles), $\rho$. Of course, the ground state energy $E$ is given by summing up on all the (relativistic) particles
\be
E(\rho)=m \int _{-B}^{B} \frac{d \theta}{2\pi} \cosh \theta \ g(\theta) \label {E3'} \, .
\ee
The free energy $f(h)$ may be defined by a Legendre transformation of the magnetic field
\be
h = \frac{dE}{d\rho}  \, ,
\ee
which realises the minimum value
\be
f(h)={ \textrm{min}}_{\rho}[E(\rho)-h\rho]
\label {fh-1'} \, .
\ee
Easily, imposing the stationarity leads to the expression
\be
f(h)=- m \int _{-B}^{B} \frac{d \theta}{2\pi} \cosh \theta \ \epsilon (\theta)    \, ,     \label {free-en}
\ee
where the 'pseudoenergy' $\epsilon (\theta )$ satisfies a linear equation with chemical potential $h$
\be
\epsilon (\theta )- \int _{-B}^{B}d\theta ' K(\theta -\theta ') \epsilon (\theta ')=h-m\cosh \theta \label {eps2'} \, ,
\ee
and the 'local' boundary condition\footnote{Due to the parity of the function $\epsilon (\theta)=\epsilon (-\theta)$, the condition $\epsilon (-B)=0$ holds as well.}
\be
\epsilon (B)=0  \, . \label {eps-B}
\ee
This condition is the local parallel of (\ref{rho3'}) and analogously determines the Fermi rapidity $B=B(h)$ as a function of the magnetic field. Its use is crucial in  determining the free energy (\ref {free-en}) as a function of $h$ (\cite{HN} and references therein).

Conversely, if we are given the free energy $f(h)$, we can compute the energy
\be
E(\rho)=f(h)+h\rho \, ,
\label  {inv-leg}
\ee
by means of the (inverse) Legendre transform
\be
\rho = - \frac{df}{dh}  \, ,
\label {dfdh}
\ee
which ought to be inverted to express $h=h(\rho)$ as a function of $\rho$.
After recalling these basic relations for relativistic models (valid strictly in the thermodynamic limit), we want to perform on them some
manipulations to obtain results usable in Section 6.

First, we differentiate twice (with respect to $\theta$) equation (\ref{eps2'})
\be
\epsilon ''(\theta )=-m \cosh \theta - [K(\theta -B)+K(\theta +B)]\epsilon '(B)+
\int _{-B}^{B}d\theta ' K(\theta -\theta ') \epsilon ''(\theta ') \, . \label {epsil2}
\ee
Then, we differentiate with respect to $B$ the equation for the density (\ref {gequat})
\be
\frac{\partial}{\partial B} g(\theta) = [K(\theta -B)+K(\theta +B)] g(B) +
\int _{-B}^{B}d\theta ' K(\theta -\theta ')\frac{\partial}{\partial B} g(\theta ') \, .
\label {derBg}
\ee
Finally, we put together the previous two (\ref {epsil2}, \ref {derBg}) and (\ref{gequat}), so that we obtain the useful mixing relation
\be
\epsilon ''(\theta ) g(B) + g(\theta) g(B) +
\epsilon '(B) \frac{\partial}{\partial B} g(\theta)=0 \, .
\label {master}
\ee
This equation is conveniently integrated from $-B$ to$B$:
\be
2 \epsilon '(B) g(B) +2\pi \rho (B) g(B) + \epsilon '(B) \int _{-B}^{B} d\theta \frac{\partial}{\partial B} g(\theta)  =0 \, . \label {master1}
\ee
The last integral appears when differentiating (\ref {rho3'}) with respect to $B$
\be
2\pi \frac{d}{dB} \rho (B)= 2 g(B) +  \int _{-B}^{B} d\theta \frac{\partial}{\partial B} g(\theta) \, , \label
{useful2}
\ee
and so it can be inserted into (\ref {master1}) to yield
\be
\epsilon '(B) \frac{d}{dB} \rho (B) + \rho (B) g(B)=0 \, .
\ee
At this final stage we may re-write the previous relation in the form of the additional ($1/R^2$) term inside (\ref {f-f0})
\be
\frac{\epsilon '(B)}{g(B)}=
- \frac{\rho (B)}{\frac{d\rho (B)}{dB} } = - \frac{d}{d(\ln \frac{\rho}{m})} B\left ( \frac{\rho}{m} \right )\, ,
\label {final-rel}
\ee
where in the last relation $\rho$ is to be computed in $B$.

\section{An alternative approach for $1/R$ contribution and beyond} \label {Bexp}
\setcounter{equation}{0}

With accuracy $1/R$ equations (\ref {chi2}, \ref {epsilon2}, \ref {j2})
read as follows
\ba
\chi (\theta)&=&  m \ R \ \cosh \theta +
\int _{-B}^{B}
d\theta ' K(\theta - \theta ') \chi (\theta ') - \nonumber \\
&-& \frac{\pi ^2}{6 \chi (B)} \frac{d}{dB}[ K(B + \theta )+K(B - \theta ) ]   \, ,
\label {chi3}
\ea
\be
\varepsilon = m \int _{-B}^{B} \frac{d\theta }{2\pi} \chi (\theta) \cosh \theta -
\frac{m(g)\pi}{6} \frac{\sinh B}{\chi (B)}
\label {epsilon3} \, ,
\ee
\be
\frac{\ell \sqrt{\lambda} R(s,g)}{2\pi}=\int _{-B}^{B} \frac{d\theta}{2\pi} \chi (\theta ) \, . \quad  \label {j3}
\ee
Perturbatively, we can expand $\chi (\theta)$, $B$ and $\varepsilon$ in powers of $1/R$,
\ba
&& \chi (\theta )= \chi ^{(-1)} (\theta ) +  \chi ^{(1)}(\theta ) + O(1/(R^3) \, , \quad B=B^{(0)}+B^{(2)}+ O(1/(R^4) \, , \nonumber \\
&& \varepsilon = \varepsilon ^{(-1)} + \varepsilon ^{(1)} + O(1/R^3) \, ,
\ea
where with the superscript $^{(r)}$ we mean the proportionality of the $r$-th order: $\chi ^{(r)}, B^{(r)}, \varepsilon ^{(r)} \propto R^{-r}$.

Now, we want to constrain $\gamma $ up to the finite size correction $1/R$, namely we easily can derive
\be
\varepsilon ^{(1)}= m \int _{-B^{(0)}}^{B^{(0)}} \frac{d\theta }{2\pi} \chi ^{(1)}(\theta) \cosh \theta
- \frac{m \pi}{6} \frac{\sinh B^{(0)}}{\chi ^{(-1)}(B^{(0)})} + \frac{m}{\pi} \cosh B^{(0)} \chi ^{(-1)} (B^{(0)}) B^{(2)}  \, . \label {eps1-1}
\ee
The function $\chi ^{(1)} (\theta )$ satisfies the linear equation
\ba
\chi ^{(1)} (\theta )&=&-\frac{\pi ^2}{6 \chi ^{(-1)}(B^{(0)})}\frac{d}{dB^{(0)}}[K(\theta +B^{(0)})+K(B^{(0)}-\theta )]
+ \nonumber \\
&+& [K(\theta -B^{(0)})+K(\theta +B^{(0)})] \chi ^{(-1)} (B^{(0)}) B^{(2)}+ \int _{-B^{(0)}}^{B^{(0)}}
d\theta ' K(\theta - \theta ') \chi ^{(1) }(\theta ') \, . \label {lineq}
\ea
The idea is now to express $\chi ^{(1)}(\theta )$ contained in (\ref {eps1-1}) in terms of the resolvent
(with respect to the kernel $K$ in the interval $[-B^{(0)}, B^{(0)}]$) of linear equation (\ref {lineq}).
Then the integration in $\theta $ generates the function $\chi ^{(-1)}$ and its first derivative
in $B^{(0)}$. Eventually we are left with
\be
\varepsilon ^{(1)}= - \frac{\pi {\chi '}^{(-1)} (B^{(0)})}{6 R \chi ^{(-1)}(B^{(0)})}
+  \frac{1}{\pi R}
[\chi ^{(-1)}(B^{(0)})]^2 B ^{(2)} \, . \label {epsi-1}
\ee
The coefficient $B ^{(2)}$ is found\footnote{We remark that - despite the use of a similar notation - the quantity $B^{(2)}$ used in this Appendix is different from the function
$B^{(2)}(h)$ given in (\ref {cond-2}). Indeed, the quantity $B$ appearing in (\ref {chi3}-\ref{j3}) is actually a function of $\rho$: therefore, it can be obtained from $B(h)$ appearing in Section 6 after expressing $h$ in terms of $\rho$ by means of the function $h(\rho)$ coming from (\ref {rho-h}): $B=B(h(\rho))$. Since $h(\rho)$ expands in powers of
$1/R^2$ too, $h(\rho)=h^{(0)}(\rho)+h^{(2)}(\rho)+...$, we have the equalities: $B^{(0)}=B^{(0)}(h^{(0)}(\rho))$, $B^{(2)}=B^{(2)}(h^{(0)}(\rho))+
\frac{\partial B^{(0)}}{\partial h^{(0)}}(h^{(0)}(\rho)) \, h^{(2)}(\rho)$.}
after writing equation (\ref {j3}) at order $1/R$, obtaining
\be
\int _{-B^{(0)}}^{B^{(0)}}\frac{d\theta}{2\pi} \chi ^{(1)} (\theta) + \frac{B ^{(2)}}{\pi} \chi ^{(-1)}(B^{(0)})=0 \, .
\ee
This relation is studied with the help of
\be
\chi ^{(1)} (\theta)=- \frac{\pi ^2}{6 [\chi ^{(-1)} (B^{(0)})]^2} \left [ {\chi ''}^{(-1)}(\theta)-
{\chi }^{(-1)}(\theta) + \frac{ {\chi '}^{(-1)} (B^{(0)})}{\chi ^{(-1)}(B^{(0)})}
\frac{\partial }{\partial B^{(0)}} \chi ^{(-1)}(\theta) \right ] + B^{(2)}\frac{\partial }{\partial B^{(0)}} \chi ^{(-1)}(\theta)
\ee
and gives
\be
B^{(2)}=\frac{\pi ^2}{6 [\chi ^{(-1)} (B^{(0)})]^3} \left [ {\chi '}^{(-1)} (B^{(0)})-{\chi }^{(-1)} (B^{(0)})
\frac{\rho (B^{(0)})}{\frac{d\rho (B^{(0)}) }{dB^{(0)}}} \right ] \, . \label {B2-fin}
\ee
Plugging (\ref {B2-fin}) into (\ref {epsi-1}), we finally obtain the desired quantity
\be
\varepsilon ^{(1)}=-\frac{\pi}{6R} \frac{\rho (B^{(0)})}{\frac{d \rho (B^{(0)}) }{dB^{(0)}}} \, .
\ee
which coincides with (\ref {first-corr}). By construction, this treatment may be generalised to higher orders.

\section{UV regime and kink equations for $O(N)$ NLSMs} \label {eq-kink}
\setcounter{equation}{0}

In order to write (\ref {first-corr}) at strong coupling, we need to know the function $B(\rho)$ in the (bulk) $O(N)$ NLSM for large total density $\rho/m \gg 1$\footnote{As in appendix \ref {useful} also in this appendix we highlight only the dependence of the various functions on $\theta$ and drop the index $^{(0)}$ we used in Section 6.}. This corresponds to the ultraviolet (UV) or massless regime, which turns out to be the large magnetic field regime $h/m \gg1$. This appendix is devoted to this regime and, in particular, to the consequent expansion of the $B(\rho)$.

First, we shall compute the UV free energy (\ref {free-en}). We shall refine and extend the idea, similar to the TBA dilogarithm trick, which has been devised in \cite{FOZ} for the 'pseudoenergy'  $\epsilon(\theta)$. In this way, though, it would not work for the theories -- for instance the NLSMs --, where the Fourier transform of the scattering kernel $\hat K(0)=1$. This failure is due to a divergence which we will disentangle suitably.

We start by the key observation concerning the following manner to write the free energy (\ref {free-en})
\be
f(h)=-m \int _{-B}^{B}  \frac{d\theta}{2\pi} e^{\theta} \epsilon (\theta) \, . \label {free-en-2}
\ee
When the integrated density $\rho$ is large, so is integral extremum $B$ (cf. (\ref{rho3'})) and then the integral in (\ref {free-en-2}) is dominated by the behaviour of $\epsilon (\theta)$ for $\theta \sim B$. In this region, according to (\ref{eps2'}) $\epsilon (\theta) \cong \epsilon_k(\theta)$, where $\epsilon _k(\theta)$ satisfies the simpler 'kink' equation
\be
\epsilon _k(\theta)- \int _{-B}^{B}d\theta ' K(\theta -\theta ') \epsilon  _k(\theta ') =h- \frac{m}{2} e^{\theta}
 \, , \label {epskink}
\ee
with the boundary condition $\epsilon _k(B)=0$. Then, in the UV regime the free energy $f(h)$ is well approximated by the UV value
\be
f_{UV}(h)=-m \int _{-B}^{B}  \frac{d\theta}{2\pi} e^{\theta} \epsilon _k(\theta) \label {f-kink} \, .
\ee
We highlight the fact that, contrary to what done in \cite {FOZ}, in (\ref {epskink}, \ref {f-kink}) we do not replace the lower extremum of integration $-B$ with $-\infty$. This difference is crucial in order to extend the results of \cite {FOZ} to NLSMs.

Now the fundamental trick comes out. Upon differentiating  (\ref{eps2'}), we may  obtain from it, as above, a 'kink' equation \footnote{If we invert the procedures by differentiating (\ref {epskink}), we obtain an integral equation whose right hand side contains -- with respect to (\ref {epskinkder}) -- the additional term $K(\theta +B) \epsilon _k (-B)$. In any case, this term can be
safely neglected - compared to $\frac{m}{2} e^{\theta}$ - in the region $\theta \sim B$ which gives the leading contribution to the integral (\ref {f-kink}).}
\be
\epsilon '_k(\theta)- \int _{-B}^{B}d\theta ' K(\theta -\theta ') \epsilon '_k(\theta ') =- \frac{m}{2} e^{\theta} \, .
\label {epskinkder}
\ee
In (\ref {f-kink}) we substitute $- \frac{m}{2} e^{\theta}$ with the above expression and obtain
\be
f_{UV}(h)=\int _{-B}^{B}  \frac{d\theta}{\pi}  \epsilon _k(\theta) \left [ \epsilon '_k(\theta)- \int _{-B}^{B}d\theta ' K(\theta -\theta ') \epsilon '_k(\theta ') \right ]
\ee
and then, using (\ref {epskink}),
\be
f_{UV}(h)=\int _{-B}^{B}  \frac{d\theta}{\pi}  \epsilon '_k(\theta) \left [h-\frac{m}{2}e^{\theta} \right ] \, .
\ee
Now, we recognise $f_{UV}(h)$ in the r.h.s., integrate by parts and use the boundary condition $\epsilon _k(B)=0$ to reach
\be
f_{UV}(h)=- \frac{h}{\pi} \epsilon _k (-B)+ \frac{m}{2\pi}e^{-B}\epsilon _k (-B)- f_{UV}(h) \, .
\nonumber
\ee
Thanks to the UV condition $m\ll h$, we are allowed to neglect the second term in the r.h.s. with $me^{-B}\sim m^2/h$ \footnote{The condition $\epsilon (B)=0$ with the rough approximation $\epsilon(\theta) \approx (h- m \cosh \theta)$ \, implies $e^{-B}\sim m/h$.} with respect to the first one (with $h$) and thus conclude finally
\be
f_{UV}(h)= - \frac{h}{2\pi} \epsilon _k (-B)
\, .   \label {f-eps}
\ee
Notice that this value becomes that in \cite{FOZ} once $-B\rightarrow -\infty$. This limit cannot be taken, though, in NLSMs, as $\epsilon _k (-B)$ diverges as we are going to show.

In fact, we want to see how $\epsilon _k (-B)$ behaves at large $B$. First,  we observe that $\epsilon '_k(\theta)$ is exponentially depressed in the interval $-B<\theta<0$, thus we expect that the approximation
\be
\epsilon _k (-B)   \cong \epsilon _k(0)
\ee
holds (at least at leading order). Now, $\epsilon _k(0)$ is a sort of plateau value (for large $B$, similarly to TBA) and hence is easily computable by approximating (\ref {epskink}) as
\be
\epsilon _k(0)\cong h+ \epsilon _k(0) \int _{-B}^{B} d\theta K(\theta)   \, , \label {eps_k(0)}
\ee
which finally implies\footnote{The first equality of (\ref {eps-0}) has to be compared with (and extends) relation $\epsilon (-\infty)=\frac{h}{1-\hat K(0)}$ of \cite {FOZ}, valid only for models in which $\hat K(0)\not= 1$.}
\be
\epsilon _k(-B) \cong \epsilon _k(0)\cong \frac{h}{1-\int _{-B}^{B} d\theta K(\theta)}=(N-2)\frac{hB}{2}  +... \quad \, . \label {eps-0}
\ee
Upon inserting (\ref {eps-0}) into (\ref  {f-eps}) we obtain the UV free energy
\be
f(h)= - (N-2)\frac{h^2}{4\pi}B(h)+...  \quad  \, . \label {free-lead}
\ee
Here $B(h)$ is still unknown, which we may fix looking at the Legendre dual, the energy. Through the same procedure as above\footnote{This easy derivation is absent in \cite{FOZ}.}, we first re-write (\ref {E3'}) without approximations as
\be
E(\rho)  = m \int _{-B}^{B} \frac{d\theta }{2\pi} e^{\theta } \ g(\theta)
\label {En3-1} \, .
\ee
Then, in the UV limit, we can approximate
\be
E(\rho)  \cong E_{UV}(\rho) = m \int _{-B}^{B} \frac{d\theta }{2\pi} e^{\theta } \ g_k(\theta)
\label {En3kink} \, ,
\ee
where $g_k(\theta)$ satisfies the 'kink equation'
\be
g _k(\theta)- \int _{-B}^{B}d\theta ' K(\theta -\theta ') g_k(\theta ') = \frac{m}{2} e^{\theta}
\label {gikink} \, .
\ee
We differentiate (\ref {gikink}) with respect to $\theta$ and obtain the equation
\be
g_k^{\prime} (\theta)+K(\theta -B)g_k (B)-
\int _{-B}^{B}d\theta ' K(\theta -\theta ') g_k^{\prime }(\theta ')=\frac{m}{2}e^{\theta} \, , \label {gikinkder}
\ee
where we have neglected the subleading term $K(\theta+B) g_k(-B)$ in the l.h.s.\footnote{We could keep it and discard the consequent extra term appearing in (\ref{Eg-kink}). This means that only for the energy we could substitute the lower bound $-B\rightarrow -\infty$, as done in \cite{FOZ} for the free energy.}. Then, we insert (\ref {gikinkder}) into
(\ref {En3kink}) and use (\ref {gikink}): we obtain
\be
E_{UV} =\frac{g_k(B)}{\pi}\left [ g_k (B)- \frac{m}{2} e^{B} \right ]
+ \int _{-B}^{B}\frac{d\theta}{\pi} g_k^{\prime} (\theta)\frac{m}{2} e^{\theta} \, ,
\ee
which, after integration by parts in the second term in the r.h.s., becomes the leading UV energy
\be
E= \frac{[g_k(B)]^2}{2\pi}+ \dots  \, .      \label {Eg-kink}
\ee
We remark that this relation actually holds for any relativistic two dimensional
integrable field theory as we did not use any feature of the $O(N)$ NLSMs.

Finally, we need two more kink relations descending from (\ref {gikink}, \ref {epskinkder}) and the easily derivable equation
\be
\frac{\partial }{\partial B} g_k(\theta)= K(\theta -B) g_k(B) + \int _{-B}^{B}d\theta ' K(\theta -\theta ') \frac{\partial }{\partial B} g_k (\theta ') \, . \nonumber
\ee
These read
\be
\epsilon ' _k(\theta) + g_k'(\theta)+ \frac{\partial }{\partial B} g_k(\theta)= 0 \, , \quad g_k(\theta)=- \epsilon '_k(\theta)  \, . \label {kink-rel}
\ee
Inserting the second of (\ref {kink-rel}) into the first one and evaluating the latter at $\theta =B$, we obtain the differential equation \be
g_k(B)=  \frac{\partial }{\partial B} g_k(B)      \, ,
\ee
which obviously implies that $g_k(B)=(const) \ e^{B}$. Thus, (\ref {Eg-kink}) yields the simple and interesting exponential dependence
\be
E_{UV}=(const.) \ e^{2B} \, .
\ee

Now we can conclude with the determination of the functions $B(h)$ and then $B(\rho)$. First of all, Legendre relations (\ref {inv-leg}, \ref {dfdh}) say that
$E=f(h)-h\frac{df}{dh}$, i.e., using  (\ref {free-lead}),
\be
E=  (N-2) \frac{h^2}{4\pi} \left [B(h)+h \frac{\partial B}{\partial h} \right ] +... \quad  \, .
\ee
Since $E=(const) \ e^{2B}\left (1+...\right )$,
we have to cope with the differential equation,
\be
(N-2) \frac{h^2}{4\pi} \left [B(h)+h \frac{\partial B}{\partial h} \right ]+...=(const) \ e^{2B}\left (1+...\right )
\, , \label {diff-eq}
\ee
which obviously holds only for the leading order.
This equation (\ref {diff-eq}) is solved by the Ansatz $B(h)=a \ln \frac{h}{m}+b \ln \ln \frac{h}{m}+...$, with $a$, $b$ constants. As seen the first term is naturally suggested by the boundary condition $\epsilon (B)=0$. Equation (\ref {diff-eq}) fixes the values $a=1$, $b=1/2$:
\be
B(h)=\ln \frac{h}{m}+\frac{1}{2}\ln \ln \frac{h}{m} +... \quad \, .  \label  {B-h}
\ee
An interesting remark may be that the first two leading terms of this UV expansion are 'universal', in the peculiar sense that they are the same for all the $O(N)$ NLSMs. Now, we need only to express $h$ in terms of $\rho$ by the Legendre transform $\rho =-\frac{df}{dh}$, which can be inverted as
\be
\ln \frac{h}{m}=\ln \frac{\rho}{m}-\ln \ln \frac{\rho}{m}+... \, .
\ee
In fact, this implies for (\ref {B-h}) that
\be
B(\rho)=\ln \frac{\rho}{m}-\frac{1}{2}\ln \ln \frac{\rho}{m} +...\label {Brhom}
\ee
and this is what we need in order to compute the Casimir energy up to two string loops. Of course, these two leading terms are universal as above.


\end{document}